\documentclass[aps, pre,twocolumn,superscriptaddress,superscriptreference]{revtex4-1}

\usepackage{graphicx}
\usepackage{dcolumn}
\usepackage[usenames,dvipsnames]{xcolor}
\usepackage[colorlinks, linkcolor=BrickRed, urlcolor=blue!50!black,citecolor=blue!50!black]{hyperref}
\usepackage{bm,mathtools}
\usepackage{natbib} 
\usepackage{tabularx} 
\usepackage{amsmath}  
\usepackage{color}
\usepackage{float}

\newcommand{\hf}{\hat{F}_0}
\newcommand{\hv}{\hat{\mathbf{v}}}
\newcommand{\hdr}{\hat{D}_R}
\newcommand{\htt}{\hat{t}}

\newcommand{\kbT}{k_\mathrm{B}T}
\newcommand{\kbTe}{k_\mathrm{B}T_{\mathrm{eff}}}
\newcommand{\kbTt}{k_\mathrm{B}\tilde{T}}
\newcommand{\tlj}{t_\mathrm{LJ}}

\definecolor{darkgreen}{rgb}{0,0.5,0}
\newcommand{\fs}[1]{\textcolor{black}{#1}}
\newcommand{\gj}[1]{\textcolor{black}{#1}}
\newcommand{\js}[1]{\textcolor{black}{#1}}

\newcommand{\gjnew}[1]{\textcolor{black}{#1}}
\newcommand{\jsnew}[1]{\textcolor{black}{#1}}
\newcommand{\fsnew}[1]{\textcolor{black}{#1}}

\newcommand{\finaledits}[1]{\textcolor{red}{#1}}

\newcommand{\comment}[1]{}

\newcommand{\fsc}[1]{\comment{\fs{#1}}}
\newcommand{\gjc}[1]{\comment{\gj{#1}}}

\begin{document}

\title{Force renormalization for probes immersed in an active bath}
\author{Jeanine Shea}
\email{j.shea@tu-berlin.de}
\affiliation{Institut f\"{u}r Theoretische Physik, Technische Universit\"{a}t Berlin, 10623 Berlin, Germany.}

\author{Gerhard Jung}
 \affiliation{Laboratoire Charles Coulomb (L2C), Universit\'{e} de Montpellier, CNRS, 34095 Montpellier, France}
 
 \author{Friederike Schmid}
 \email{friederike.schmid@uni-mainz.de}
 \affiliation{Institut f\"{u}r Physik, Johannes Gutenberg-Universit\"{a}t Mainz, 55099 Mainz, Germany}

\begin{abstract}
Langevin equations or generalized Langevin equations (GLEs) are popular models for describing the motion of a particle in a fluid medium in an effective manner.  Here we examine particles immersed in an inherently nonequilibrium fluid, i.e., an active bath, which are subject to an external force. Specifically, we consider two types of forces that are highly relevant for microrheological studies: A harmonic, trapping force  and a constant, ``drag'' force.  We study such systems by molecular simulations and use the simulation data to derive an effective GLE description. We find that, in an active bath, the external force in the GLE is not equal to the physical external force, but rather a renormalized external force, which can be significantly smaller. The effect cannot be attributed to the mere temperature renormalization, which is also observed.
\end{abstract}

\maketitle

\section{Introduction}
\label{sec:mk}
Microrheology focuses on linking the microscopic dynamics of a system to its macroscopic response. The goal of microrheological studies is then to investigate the behavior of an immersed and experimentally detectable probe particle to gain knowledge about the surrounding medium which is \js{made up of smaller particles}~\cite{microrheology1,microrheology2,microrheology3}. \js{Microrheological techniques have been used to study a variety of different mediums ranging from complex fluids~\cite{microrheology_dna,microrheology_complexfluid,micro_complexfluid2,micro_complex3} to granular matter~\cite{Zik_1992,dauchot_granular1,dauchot_granular2,seguin_granular,Fiege_2012_granular}.}

In general, microrheological studies can be classified in two categories: passive and active~\cite{microrheology4,microrheology5}. Passive microrheology monitors the diffusive motion of an immersed probe to extract information about fluctuations and transport properties of the medium. Active microrheology relies on subjecting a probe to an external force to determine the properties of the immersive medium. This is true for both experimental and computational microrheological studies. The external forces often take the form of either constant, drag forces \cite{Seifert_2010,cordoba2012elimination,wip} or harmonic, trapping forces \cite{Daldrop2017_external}. Microrheological studies employing these forms of external forces have also recently been applied to probe the properties of active systems~\cite{Chen,Stark_Milos,Harmonic_Drag_Brady,Seyforth,Burkholder_Brady,Burkholder_Brady2}.

Since microrheology focusses on investigating a small number of immersed probes, usually described by their positions and velocities, the extracted information will correspond to a strongly coarse-grained description, in which the surrounding medium is replaced by effective friction coefficients and fluctuations.  In previous studies, for example, the system of a passive probe immersed in an active bath and subject to a harmonic potential has been mapped onto an overdamped Langevin equation~\cite{From_RohanJ,maggi,Jayaram_2023}. However, in other studies, it has been shown that, under certain conditions, the dynamics of a probe immersed in an active bath cannot be assumed to be Markovian~\cite{Voigtmann,abbasi2022non}. In this case, the Langevin equation is no longer sufficient and, instead, a generalized Langevin equation (GLE) is required to capture the system dynamics in a coarse-grained model~\cite{Zwanzig_Non,memory_review}.

\fs{A simple and popular Ansatz for such a GLE is to assume the form \cite{Izvekov2013_PO,memory_review,proj_op,wip,Izvekov2013_PO_NEQ,Vroylandt_2022_FDT,netz2023derivation}}
 \begin{equation}
\label{eq:gle}
M \dot{\mathbf{V}}(t)= \mathbf{f}(\mathbf{R},t)-\int^t_{\fsnew{T}} \mathrm{d}s \: \mathbf{K}(t-s) \mathbf{V}(s) 
+  \bm{\Gamma}(t),
\end{equation}
where $M$ is the particle mass, $\mathbf{V}(t)$ its velocity, $\mathbf{R}(t)$ its position, $\mathbf{f}(\mathbf{R},t)$ describes the drift induced by the external forces (in passive microrheology, $\mathbf{f} \equiv 0$), $\mathbf{K}(t-s)$ is a memory kernel, and $\bm{\Gamma}(t)$ a colored stochastic noise with zero mean. \gj{Vroylandt and Monmarch\'{e} \cite{Vroylandt_2022_FDT, Vroylandt_2022} have recently shown that it is possible to derive this form of the GLE -- possibly with a space-dependent memory kernel -- from first principles by projection operator techniques \cite{MZ, Zwanzig_Non} for Hamiltonian systems at equilibrium.} \fs{In the absence of external forces (at $\mathbf{f}(\mathbf{R},t) = 0$), the noise term in the GLE as obtained from the projection operator
 obeys a so-called second fluctuation dissipation theorem (2FDT)
 $\langle \bm{\Gamma}(t_0) \bm{\Gamma}(t) \rangle_{\mathrm{eq}}=\kbT \mathbf{K}(t-t_0)$}.
\fs{The effect of including external forces in} a GLE description\cite{KIM1972469,10.1063/5.0159283,netz2023derivation,koch2023nonequilibrium}, non-stationary dynamics \cite{Meyer2017,Schilling_2022}, and stochastic microscopic dynamics \cite{Zhu_2022,Zhu_2023} has also been discussed using first principle theory. Important conclusions from these studies are that \jsnew{the 2FDT is no longer generally valid} \fsnew{in such cases}: namely, in the presence of non-linear or stochastic interactions \cite{Glatzel_2021,Zhu_2023}, external forces, or general non-equilibrium settings \cite{Jung_2022}. Furthermore, additional terms in this relation \fsnew{can} \jsnew{emerge} \cite{Vroylandt_2022_FDT, netz2023derivation,Zhu_2023,10.1063/5.0049324}.

\gj{ In this work we will take a complementary approach and use the GLE (\ref{eq:gle}) as a starting point to extract effective equations of motion for the coarse-grained variables from microscopic simulations.} \fs{Thus, our goal is not
to {\em derive} a GLE model (\ref{eq:gle}) from a microscopic Hamiltonian (which may not be rigorously possible), but to
{\em map} an ensemble of microscopic trajectories onto the model (\ref{eq:gle}) such that important static properties, such as the distribution of particle positions, and important dynamic properties, such as the average drift and velocity correlation functions, are reproduced by the dynamical system defined by the GLE.} \gjnew{All averages, $\langle...\rangle$, will therefore refer to the average over this, possibly non-equilibrium, ensemble.} 
One should note that one has some freedom as to how to distribute the total force between $\mathbf{f}$, $\mathbf{K}$ and $\bm{\Gamma}$ \cite{Zwanzig_Non,Mitterwallner_2020,wip}. At equilibrium, a popular Ansatz \cite{proj_op,doi:10.1021/acs.jctc.2c00871,10.1063/5.0049324,10.1063/5.0163097} is to assume that $\mathbf{f}$ subsumes all reversible interactions (which also include medium-mediated interactions) and therefore correspond to the force emerging from static coarse-graining. The other two terms then characterize the thermal coupling with the medium; i.e., they can be seen as a thermostat \cite{ceriotti2010colored} with colored noise. 

What happens in active systems far from equilibrium? This question has been intensely discussed in the literature \cite{Chen,Seyforth, maggi_fdt, Maes2014,Burkholder_Brady,Steffenoni2016,Doerries_2021,Jung_2022,baldovin2022many,Mitterwallner_2020}, often focusing on the question of whether the 2FDT is still fulfilled. However, as mentioned above, one has some freedom as to how to distribute the net bath forces between the different terms of the GLE. In the absence of drift forces, it has been shown \cite{Shin_2010, wip} that many memory reconstruction methods \cite{Shin_2010,Carof_2014,Jung_2017,Daldrop2017_external, GLD,Bockius_21} will yield GLEs that automatically satisfy a generalized 2FDT, 
\begin{equation}
\label{eq:2fdt_gen}
 \fs{\langle \bm{\Gamma}(t_0) \bm{\Gamma}(t) \rangle 
= M \mathbf{K}(t-t_0) \: \langle \mathbf{V}\mathbf{V} \rangle}  
\end{equation}
(in tensorial notation). Putting it differently: \gj{One can impose an orthogonality condition between the coarse-grained variables and the stochastic noise $\langle \bm{V}(0) \bm{\Gamma}(t) \rangle = 0$\fs{, where $t=0$ refers to some arbitrary ''initial time'',}  as an additional condition on the memory kernel of the force-free GLE}. Mathematically, \gj{this implies that  the memory kernel and the velocity autocorrelation function depend} on each other via a so-called Volterra equation \cite{Shin_2010, memory_review}. \gj{This imposed constraint is reminiscent of the typical projector used in projection operator techniques, which would lead to similar orthogonality conditions  \cite{Zwanzig_Non,Vroylandt_2022,Zhu_2023,netz2023derivation}. We will impose this condition throughout the manuscript for all our analytical and numerical analysis.} \js{However, we would like to emphasize that we do not apply any projection operator techniques, but rather solely apply this orthogonality condition.}

In the present paper, we then address the following question: How does the active bath affect the drift force $\mathbf{f}$ on the probe in the presence of external drag forces? 
%
%
%
%
As prototype test systems, we consider a passive probe in a homogeneous bath of active Langevin particles~\cite{ALPs_Takatori,
Hidden_entropy_Marchetti,ME} subject to two typical types of external forces $\mathbf{F}_{\mathrm{ext}}$: A constant drag force and a harmonic trap force.

Our main result can be summarized as follows: We can obtain a consistent GLE mapping of this system, but only if we renormalize the force: The drift force differs from the applied force by a factor $\alpha$ which is much smaller than one. The factor does not depend on \js{whether the potential is linear or harmonic}. Physically, this means that the active bath particles help the probe particle to access high potential regions not only by increasing its kinetic energy (i.e., effectively increasing its temperature), but also by actively pushing it into these regions.

By a mathematical analysis, we identify a generalized equipartition theorem which characterizes the relation between the renormalized force and the kinetic energy of the particle. \gj{The analysis also shows that the generalized 2FDT in our model may acquire an extra term in non-equilibrium systems, consistent with Refs.\ \cite{Maes2014,Steffenoni2016,netz2023derivation}}, 
which however vanishes within the error in the systems considered here.


Our results are of high importance for the interpretation of active microrheology in active systems, as they show how to extract information about \js{an active bath by observing transport of immersed tracer particles. Our results are particularly pertinent to systems in which the constituents are of non-negligible mass, i.e. where the constituents are meso- or macroscopic. Examples of such systems include synthetic active robots~\cite{Soudeh_Lowen,leoni_robots,Tapia-Ignacio_2021} and granular active matter~\cite{des_granular3,weber_granular2,Narayan_granular1,kudrolli_granular4}. Whereas there exist previous studies on the microrheological properties of active granular matter~\cite{Zik_1992,dauchot_granular1,dauchot_granular2,seguin_granular,Fiege_2012_granular}, synthetic active robots is an emerging field and we are not aware of any studies of microrheological studies of such systems.}

\section{Model and Simulation Details}
\label{sec:syssim}

We consider a three dimensional system of a passive probe immersed in
a bath of active Langevin particles (ALPs) of mass $m$ and radius $R$,
which propel themselves with a constant force $F_0$ subject to
rotational diffusion with a diffusion constant
$D_R$~\cite{ABP_ALP,ALPs_Takatori,
Inertial_delay,Hidden_entropy_Marchetti,Enhanced_diff_Marchetti,MITD,Gompper_local_stress,Loewen_TD_inertia,ME}. \js{ An analogous model in the overdamped limit (a probe immersed in a bath of active Brownian particles with mass $m \to 0$) has often been focused on in previous studies~\cite{Ash,Brady_curved,Brady,steff,Stenhammar_MIPS,Redner_MIPS,From_RohanJ}. These ABP fluids have a number of similar properties to ALP fluids, including the quadratic scaling of the effective temperature (or effective diffusion constant in the case of an ABP) with the bath activity and the velocity correlations~\cite{ME}.} 

The ALPs are coupled to a thermal bath with temperature $\kbT$ via a Markovian, Langevin thermostat, and they interact with each other and with the immersed probe by repulsive hard core interactions of the Weeks-Chandler-Anderson (WCA) type
\cite{WCA}. Specifically, we follow Refs.~\cite{ALPs_Takatori,
Hidden_entropy_Marchetti,ME} and set the rotational inertia of ALPs to
zero for simplicity.  The equations of motion for an ALP, $n$, in the
bath, are thus given by
\begin{equation}
\label{eq:alp_int}
\begin{split}
m \dot{\mathbf{v}}_n(t)&=F_0\mathbf{e}_n(t)-\gamma \mathbf{v}_n(t)+\boldsymbol{\xi}_n(t)\\
&-\nabla U_\mathrm{WCA}(\mathbf{r}_n-\mathbf{R})-\sum_{n\neq m} \nabla U_\mathrm{WCA}(\mathbf{r}_n-\mathbf{r}_m),
\end{split}
\end{equation}
where $F_0$ is the propulsion force of the active particle,
$\mathbf{e}(t)$ is the orientation of the active particle, and
$\gamma=6\pi\eta R$ is the damping constant for an ALP radius $R$ in a
thermal bath with viscosity $\eta$. The term $\boldsymbol{\xi}_n(t)$
represents a stochastic force that mimics implicit collisions of the
ALPs with thermal bath particles. These collisions are modelled as
Gaussian white noise with mean zero and variance given by
a fluctuation-dissipation relation
\begin{equation}
\label{eq:trans_diff}
\langle \xi_i(t)\xi_j(t') \rangle = 2 \gamma \kbT \delta_{ij} \delta (t-t').
\end{equation}
The resulting translational diffusion coefficient of isolated ALPs
is given by $D_T=\kbT/\gamma$. 
Finally, the terms $-\nabla U_\mathrm{WCA}(\mathbf{r}_n-\mathbf{R})$
and $-\sum_{n\neq m} \nabla U_\mathrm{WCA}(\mathbf{r}_n-\mathbf{r}_m)$
in Eq.\ \eqref{eq:alp_int} describe the WCA interactions with the probe
and with all other ALPs, respectively. 

The time evolution of the orientation of the ALP, $\mathbf{e}(t)$, is
governed by rotational diffusion, 
\begin{equation}
\label{eq:rot}
\dot{\mathbf{e}}(t)=\mathbf{N}(t) \times \mathbf{e}(t),
\end{equation}
where $\mathbf{N}(t)$ is again Gaussian white noise with a mean of zero 
and a variance (another fluctuation-dissipation relation)
\begin{equation}
\label{eq:rot_diff}
\langle N_{\alpha}(t)N_{\beta}(t') \rangle = 2 D_R \delta_{\alpha \beta} \delta (t-t').
\end{equation}
Here $D_R$ is the rotational diffusion constant, which is given by
$D_R=3D_T/4R^2$ for a particle of radius $R$. The immersed probe only experiences the externally applied force and forces from interactions with the surrounding ALPs. Unlike the ALPs, it is not coupled to the
thermal bath. \js{Ref.~\cite{ME} has shown that the behavior of a probe immersed in an ALP fluid remains qualitatively the same regardless of whether it is coupled to the thermal bath. In particular, the probe acquires an enhanced kinetic temperature which scales quadratically with the activity of the ALP fluid even in the presence of a thermostat.}

All simulations are performed using LAMMPS \cite{lammps}. The length, energy,
and mass scales in the simulation system are defined by the Lennard-Jones (LJ)
diameter $\sigma$, energy $\epsilon$, and ALP mass $m$, respectively,
which defines the LJ time scale $\tlj = \sigma
\sqrt{m/\epsilon}$. We use truncated and shifted LJ potentials with
the energy scale $\epsilon$ which are cut off at
$r_{\mathrm{c}}=2^{\frac{1}{6}}\sigma$, resulting in purely repulsive
WCA interactions as described above. The cubic simulation box has
periodic boundary conditions in all three dimensions and a side length
based on the desired density of the bath. The probe has a mass of $M=100m$ and is defined as a
rigid body with a radius $R_p=3\sigma$. The body of the probe is
constructed so that its surface is smooth, resulting in full slip
boundary conditions for the LJ fluid. The active bath consists of ALPs
with a mass of $m_{\mathrm{ALP}}=1m$ and a radius of $R=0.5\sigma$.
The number of ALPs in the bath is determined by the desired
density of the bath. The parameters of the thermal bath are chosen
such that $\kbT = 1 \; \epsilon$ and 
$\eta = 1 \; m/(\sigma \tlj)$, resulting in
$\gamma = 3 \pi m/\tlj$, $D_T = (3 \pi)^{-1} \sigma^2/\tlj$,
and $D_R = \pi^{-1} /\tlj$.
\js{For the active bath, we choose an active force of $F_0 = 50 \: m \sigma^2/t_{lJ}$ because Ref.~\cite{ME} showed that this force is sufficiently large to induce non-equilibrium probe dynamics. but not large enough to cause motility induced phase separation in the active bath.}

In the following, for consistency with previous work \cite{ME}, we will often also use dimensionless quantities $\htt = t \: \gamma/m$, $\hv = \mathbf{v} \: \sqrt{{m}/{\kbT}}$, $\hf = F_0 \frac{1}{\gamma} \sqrt{{m}/{\kbT}}$, and $\hdr = D_R \: m/\gamma$. Dimensionless quantities will denoted with a hat. Density values and the spring constant of the harmonic potential, and all distances are given in LJ units. In these cases, the LJ units are explicitly written with the value. \fsc{I removed the comment on values without dimensions being dimensionless, because we may still have missed some in the paper.}

\section{Results}

The behavior of free particles in the ALP bath has been reported in previous work \cite{ME} for different bath densities and active forces $F_0$. One important result of this study was that the bath transmits a kinetic temperature to the probe ($\kbTe = M \langle \mathbf{V}^2 \rangle/3$ in 3D), which, somewhat unexpectedly, can be significantly higher than the effective kinetic temperature of the ALP bath particles. In the present article, we will analyze the probe dynamics in the presence of external forces as described below.

\subsection{Harmonic potential}
\label{sec:harm}
We trap the immersed probe in a harmonic potential such that its equation of motion is:
\begin{equation}
\label{eq:eom_coll_harm}
M \dot{\mathbf{V}}(t)=-\nabla U_\mathrm{Harm}(|\mathbf{R}-\mathbf{R}_0|)-\sum_n \nabla U_\mathrm{WCA}(\mathbf{R}-\mathbf{r}_n),
\end{equation}
where $M$ is the mass of the probe, $\mathbf{V}(t)$ is its velocity, and $U_\mathrm{WCA}(\mathbf{R}-\mathbf{r}_n)$ is the WCA potential due to an ALP particle, $n$, at position $\mathbf{r}_n$. $U_\mathrm{Harm}(|\mathbf{R}- \mathbf{R}_0|)=\frac{1}{2}k \: (\mathbf{R}-\mathbf{R}_0)^2$ is the harmonic, trapping potential with trapping constant $k$ and trap center $\mathbf{R}_0$. We test three different values of $k$: $k=1,~5,$ and $10$ in LJ units of $\epsilon\sigma^{-2}$.

We map our system dynamics onto the GLE \cite{Daldrop2017_external,memory_review}:

\begin{equation}
\label{eq:gle_harm}
M \dot{\mathbf{V}}(t)=-\nabla U_\mathrm{Harm}^{\mathrm{eff}} (|\mathbf{R}-\mathbf{R}_0|)
-\int^t_{\fsnew{- \infty}} \mathrm{d}s \: K(t-s) \mathbf{V}(s)+\bm{\Gamma}(t).
\end{equation}
where $M$ is the particle mass, $\mathbf{V}(t)$ is its velocity, $K(t-s)$ is the memory kernel, $\bm{\Gamma}(t)$ is the stochastic force, and we allow for the possibility that $U_\mathrm{Harm}^\mathrm{eff} = \frac{1}{2} \tilde{k} \mathbf{R}^2$ may differ from $U_\mathrm{Harm}$, as is known also for non-linear systems \cite{Zwanzig_Non,jungjung2023}. 
\fsnew{Here and throughout, we will use the stationary GLE \cite{Shin_2010}, 
i.e., the initial time $T$ in Eq. (\ref{eq:gle}) is taken to be in the infinite past.}
Following a strategy outlined in Ref.~\cite{Shin_2010}
(see also appendix \ref{sec:app_memory_reconstruction}), we use Eq.~\eqref{eq:gle_harm} to construct a Volterra equation of the second kind,

\begin{equation}
\label{eq:volterra2_2}
C_F(t) = M \tilde{k} \:C_V(t) + \int_{\jsnew{-\infty}}^t \mathrm{d}s \: K(s) C_{FV}(t-s) + M K(t) C_V(0)
\end{equation}
which relates the memory kernel with the force autocorrelation function $C_F(t) = \langle \mathbf{F}(t) \cdot \mathbf{F}(0) \rangle$ (with $\mathbf{F}= M \dot{\mathbf{V}}$), the velocity autocorrelation function (VACF) $C_V(t) = \langle \mathbf{V}(t) \cdot \mathbf{V}(0) \rangle$, and the cross correlation  $C_{FV}(t) = \langle \mathbf{F}(t) \cdot \mathbf{V}(0) \rangle$. Eq. (\ref{eq:volterra2_2}) can then be used
to determine $K(t)$ following a modification of an algorithm presented in Ref. \cite{Shin_2010}. Details are shown in appendix \ref{sec:app_memory_reconstruction}. 

The corresponding 2FDT \gj{for GLE (\ref{eq:gle_harm})} in the presence of a harmonic potential
is derived in \js{Appendix} \ref{sec:app_mem}: 
\begin{equation}
    \label{eq:2fdt_mod}
    \langle \bm{\Gamma}(0) \bm{\Gamma}(t) \rangle
     = M K(t) \: \langle \mathbf{V} \mathbf{V} \rangle 
       + \tilde{k} \: I(t) \langle \mathbf{V} \mathbf{R} \rangle  
\end{equation}
with $I(t) = \int_{t}^{\infty} \textrm{d}s \: K(s)$. It differs from
Eq.\ (\ref{eq:2fdt_gen}) by an additional term which is proportional to the 
instantaneous correlations between the position and the velocity. 
This term is zero at equilibrium due to time reversal
symmetry, but could, in principle, be relevant in stationary active systems.
In the systems considered in the present work, however, 
$\langle \mathbf{V} \mathbf{R} \rangle$ was always found to be zero within
the statistical error (data not shown).  We infer that the additional term 
can be neglected and the coarse-grained GLE for our systems satisfies 
the usual generalized 2FDT,  Eq.\ \eqref{eq:2fdt_gen}.

\subsubsection{Position probability distribution}
\label{sec:harm_pos}
\begin{figure}
  \centering
  \includegraphics[width=.9\linewidth]{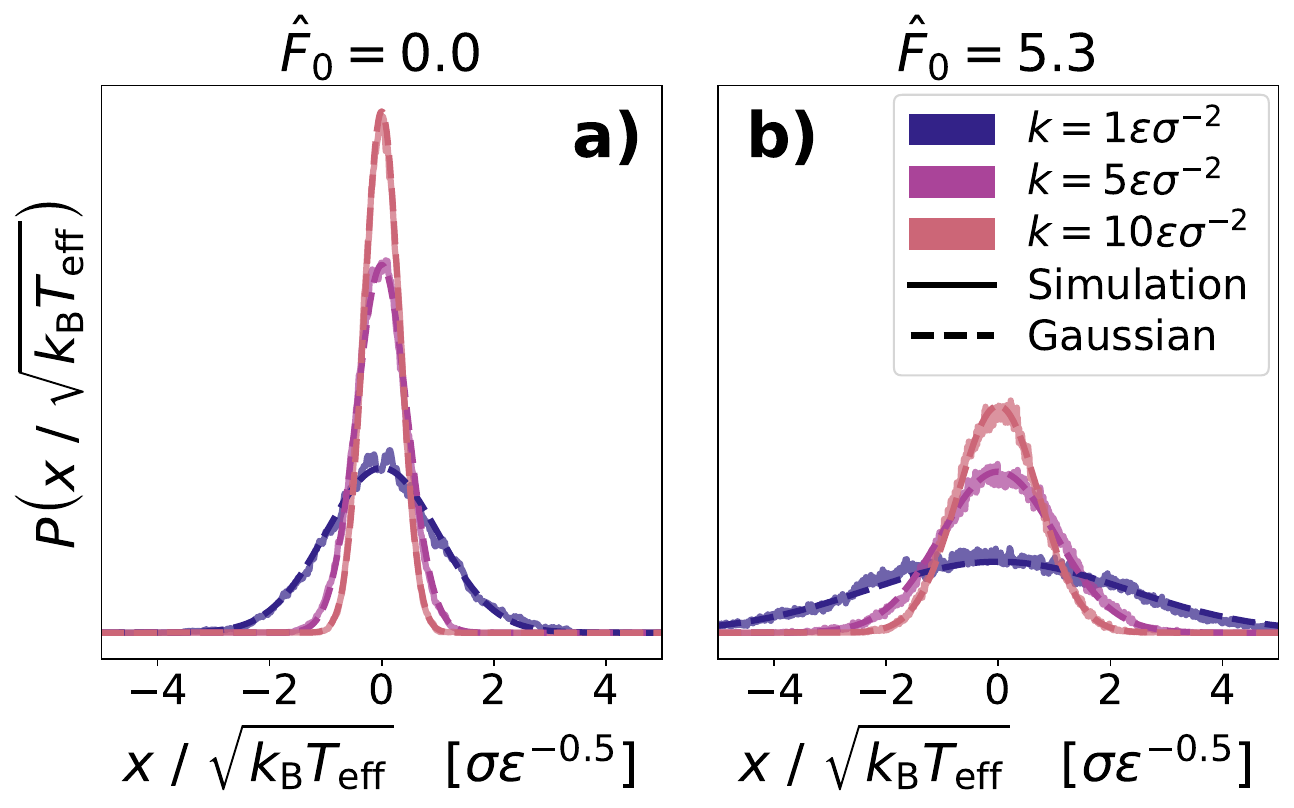}  
\caption{Position distribution in x-direction ($P(x)$) relative to the trap center of a harmonically trapped probe immersed in \textbf{a)} a passive ($\hf=0.0$) and \textbf{b)} an active bath ($\hf=5.3$). Solid lines show simulation data, whereas dashed lines show a Gaussian distribution with mean $0$ and the variance \textbf{a)} expected for a probe in a harmonic potential with each respective spring constant $k$ and \textbf{b)} of the simulation data. Each bath has average density $\rho=0.4\sigma^{-3}$. Values of $k$ are shown in the legend.}
\label{fig:harm_pos}
\end{figure}

In equilibrium, a harmonically trapped particle has a Gaussian position distribution centered at the trap center with variance $\langle (R_i-R_{0,i})^2\rangle=\kbT/k$ for each  axis $i$. As expected, this matches what we find for a probe immersed in a passive bath, as we can see in Fig.~\ref{fig:harm_pos}a), which shows the position distribution of the probe in the x-direction relative to the trap center. Although we only show the position distribution in the x-direction, since the probe is confined along all axes, we find the same form of position distribution in both the y- and z-directions as well.

We see in Fig.~\ref{fig:harm_pos}b), that the position distribution of the probe remains Gaussian and centered around the trap center even in an active bath, as in equilibrium. However, the variance of the probe position distribution is higher than \js{that of a probe immersed in a passive bath. These results are consistent with previous studies of a harmonically confined probe immersed in an active bath~\cite{Maggi_k,Jayaram_k,Volpe_k}. We note that, for higher values of $k$ when the range of confinement is comparable to the persistence length of the active particles, the distribution has been shown to become non-Gaussian~\cite{Volpe_k}. However, all our values of $k$ remain within this limit.}

\begin{figure}
  \centering
  \includegraphics[width=.8\linewidth]{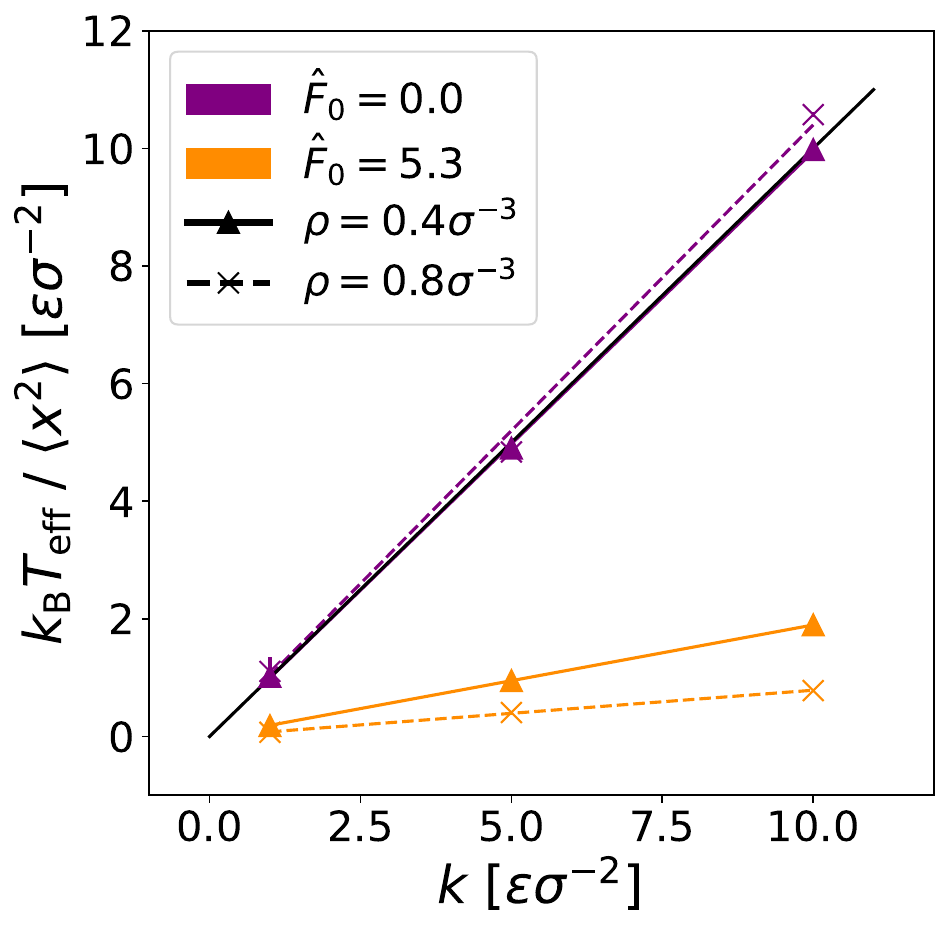}  
\caption{$\kbTe/\langle x^2\rangle$ as a function of $k$ for a probe immersed in passive baths ($\hat{F}_0=0.0$, indigo) and active baths ($\hat{F}_0=5.3$, orange) with  density  $\rho=0.4\sigma^{-3}$ (triangles) and  $\rho=0.8\sigma^{-3}$ (x-symbol). \js{Error bars on the data points are plotted, but are too small to be visible}. Solid colored lines show the line of best fit for baths with density $\rho=0.4\sigma^{-3}$, whereas dashed colored lines show the line of best fit for baths with density $\rho=0.8\sigma^{-3}$. The plot also contains a solid black line showing $\kbTe/\langle x^2\rangle=k$.} 
\label{fig:harm_x2}
\end{figure}

\js{These previous studies have attributed the enhanced variance of the position distribution to an enhanced temperature. However, when we use the definition of the enhanced kinetic temperature as quoted earlier from Ref.~\cite{ME}, we find that} $\kbT_{\mathrm{eff}}/\langle x^2 \rangle\neq k$, as can be seen in Fig.~\ref{fig:harm_x2} where we plot $\kbT_{\mathrm{eff}}/\langle x^2\rangle$ as a function of $k$. Furthermore, we see in Fig.~\ref{fig:harm_x2} that the value of $\kbT_{\mathrm{eff}}/\langle x^2\rangle$ for a given constraint $k$ is not density independent in the active case, as we would expect in equilibrium. 

Nevertheless, it does appear that $\kbT_{\mathrm{eff}}/\langle x^2\rangle\propto k$. We therefore posit that, due to the activity of the bath particles, the external force on the probe needs to be \emph{renormalized} in our coarse-grained GLE model. Then, $\kbT_{\mathrm{eff}}/\langle x^2\rangle=\Tilde{k}$ for the active bath, where $\Tilde{k}=\alpha k$ is the renormalized spring constant of the harmonic potential with renormalization factor $\alpha$.

To determine the renormalization factor $\alpha$, we fit our data in Fig.~\ref{fig:harm_x2} to a line with zero offset. As expected, in the passive case, $\alpha\approx1$ for both bath densities, i.e., the external force does not need to be renormalized and is density independent. In the active case, we find that for a bath with density $\rho=0.4\sigma^{-3}$, $\alpha_{0.4}=0.190\pm0.001$ and for a bath with density $\rho=0.8\sigma^{-3}$, $\alpha_{0.8}=0.079\pm0.002$. In both cases, $\alpha<1$, meaning that the activity of the bath effectively decreases the spring constant of the harmonic potential and, thereby, decreases the trapping force exerted on the probe. \js{In Appendix~\ref{app:alpha_f0}, we show how $\alpha$ varies as a function of $\hat{F}_0$. We find that $\alpha$ decreases from the equilibrium value $\alpha=1$ as a function of $\hat{F}_0$. Furthermore, we find that a renormalization factor is necessary even for very small bath activities ($\hat{F}_0=1$).}

\subsubsection{Velocity autocorrelation function and memory kernel}
\label{sec:harm_mk}
We now examine the velocity autocorrelation function (VACF) and memory kernel of the trapped probe. In calculating the memory kernel, we use the renormalized spring constant $\tilde{k}$ obtained from Section~\ref{sec:harm_pos}. Interestingly, we find that the reconstructed memory kernel fails to 
go to zero in the long time limit if we insert the bare spring constant or omit the spring potential altogether. In Appendix~\ref{sec:app_mem}, we analyze this problem more generally. We show that, 
for {\em any} confining renormalized external potential $U^{\textrm{eff}}$, inverting
the Volterra equation to \gj{extract} the corresponding memory kernel is bound to fail unless the following 
(tensorial) condition is fulfilled:
\begin{equation}
\label{eq:equipartition}
\langle \mathbf{R} \: \nabla U^{\textrm{eff}} \rangle = M \langle \textbf{V} \textbf{V} \rangle. 
\end{equation}
Eq. \eqref{eq:equipartition} can be seen \js{as} a generalized version of the equipartition theorem in nonequilibrium
systems. If condition \eqref{eq:equipartition} is not met, the reconstructed memory kernel will not converge to a zero value in the limit $t \to \infty$. Equation \eqref{eq:equipartition} 
is another way to characterize the renormalized potential.

\begin{figure}
  \centering
  \includegraphics[width=1.\linewidth]{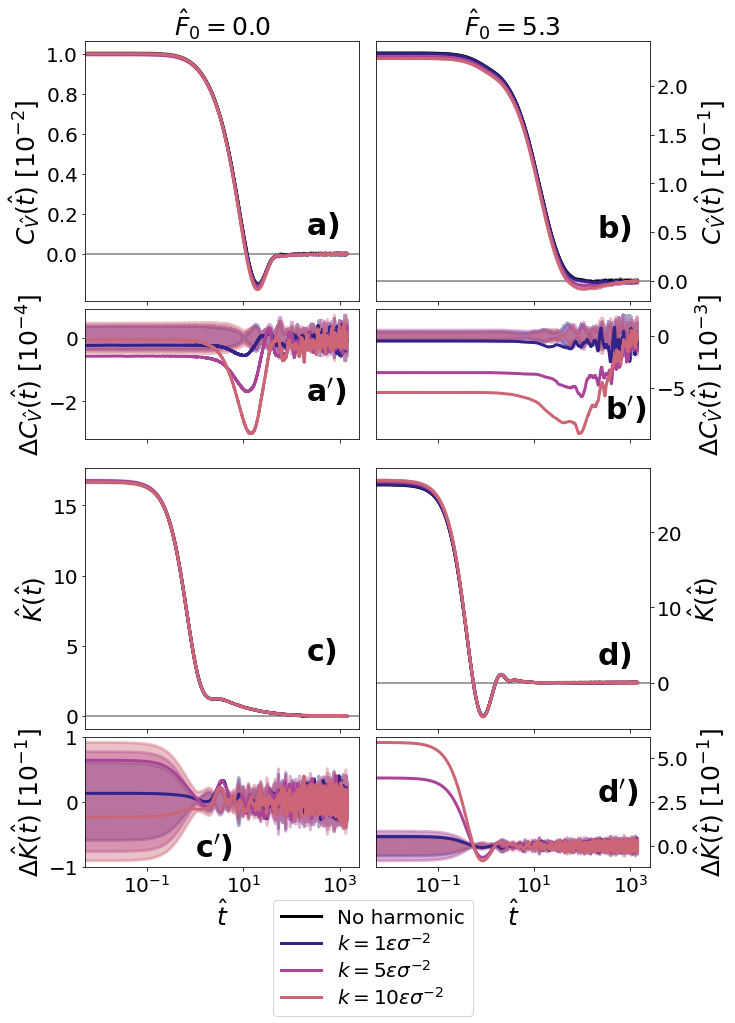}  
\caption{Velocity autocorrelation function (VACF) and memory kernel of an immersed probe trapped in a harmonic potential. The \js{larger graphs in the left column show} \textbf{a)} the VACF and \textbf{c)} the memory kernel for a probe immersed in a passive bath ($\hat{F}_0=0.0$), whereas the \js{larger graphs in the right column show} the \textbf{b)} VACF and \textbf{d)} memory kernel for a probe immersed in an active bath ($\hat{F}_0=5.3$). Light shading indicates error bars. Each bath has average density $\rho=0.4\sigma^{-3}$ and the probe has a radius $R_p=3.0\sigma$. \js{Smaller graphs correspond to the larger graph directly above them} and show the difference between the VACF/memory kernel of a free and trapped probe.}
\label{fig:harm_vv}
\end{figure}

From Figs.~\ref{fig:harm_vv}a) and c), we see that, when immersed in a passive bath, the VACF/memory kernel of a harmonically constrained probe only differs slightly from that of an unconstrained probe. Namely, the oscillation of the VACF for a harmonically constrained probe is slightly deeper than that for an unconstrained probe. The increased depth of this oscillation relates to the restoring force on the probe; therefore, we expect that its depth would increase with an increased value of $k$, 
which agrees with what we see in Fig.~\ref{fig:harm_vv}a). The memory kernel of a free vs. trapped probe in a passive bath exhibits even fewer differences. In fact, the differences between the memory kernel of a free vs. trapped particle all fall within our error bounds.

When immersed in an active bath, the VACF of the harmonically constrained probe depends more significantly on the confinement strength than in the passive case. In the inlay of Fig.~\ref{fig:harm_vv}b), it can be observed that the trapped probe has a decreased local kinetic temperature, $C_V(0)$, in comparison with that of a free probe. Furthermore, the local kinetic temperature decreases as the trapping strength $k$ increases. Within the range of $k$ values we studied, however, this effect is relatively small (two orders of magnitude smaller than that of the kinetic temperature) and cannot be readily seen in the probe velocity distributions (see Appendix~\ref{sec:app_veldist}). Therefore, this change in the kinetic temperature cannot account for the necessary rescaling of $k$.
The memory also slightly depends on $k$ at small times, but the differences are again smaller than the
error at times $\hat{t} > 0.1$.


\subsubsection{Stochastic force distribution}
\label{sec:harm_sf}

Once we have determined the memory kernel of the probe, it is straightforward to determine the stochastic force $\bm{\Gamma}(t)$: 

\begin{equation}
\label{eq:sf_harm}
\bm{\Gamma}(t)=\mathbf{F}(t)+
\nabla U_\mathrm{Harm}^{\mathrm{eff}}(|\mathbf{R}-\mathbf{R}_0|)+\int^t_{\fsnew{-\infty}} \mathrm{d}s \: K(t-s) \mathbf{V}(s).
\end{equation}

\begin{figure}
  \centering
  \includegraphics[width=\linewidth]{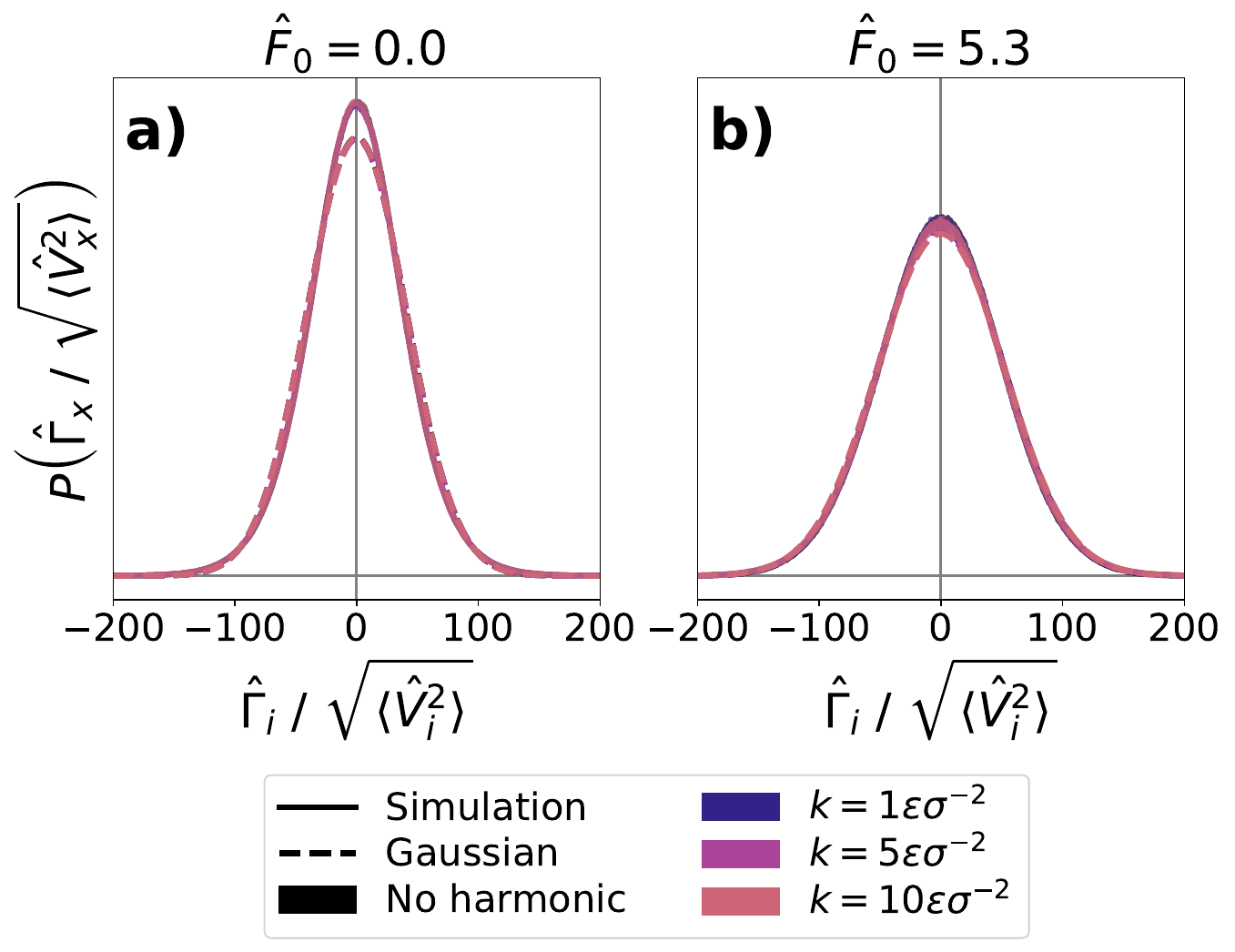}  
\caption{Stochastic force distribution of the harmonically trapped probe in a bath of density $\rho=0.4\sigma^{-3}$. The solid lines show simulation data, whereas the dotted lines show zero-centered Gaussian distributions with the same standard deviation. \textbf{a)} The stochastic force distribution for a passive bath. \textbf{b)} The stochastic force distribution for an active bath  with activity $\hat{F}_0=5.3$. In \textbf{b)}, the Gaussian distributions overlap the simulation data.}
\label{fig:harm_sf}
\end{figure}

In Fig.~\ref{fig:harm_sf}, we see that in both the case of a passive and an active bath, the stochastic force distribution calculated from simulation data (solid lines) approximately matches a zero-centered Gaussian with the same standard deviation (dashed lines). The only deviations from the zero-centered Gaussian are an enhanced peak at the center of the distribution, which we infer results from the low density of the bath and which was also seen for an unconstrained probe in Ref.~\cite{ME}. 

\subsubsection{Characterization of active bath}
\label{sec:harm_sh}
We now examine the properties of the ALP fluid in the vicinity of the probe. To this end, we analyze the angle dependent
density distribution $\rho(\mathbf{r})$ of ALPs around the probe in a comoving and corotating frame with origin
at the probe center, $\mathbf{R}(t)$, and $z$ axis always aligned in the direction of the instantantenous probe velocity $\mathbf{V}(t)$. To quantify the angular dependence, we expand the density distribution $\rho(\mathbf{r})$ 
in spherical harmonics,
\begin{equation}
 \rho(\mathbf{r}) = \sum_{lm} Y_{lm}^*(\mathbf{r}/r) \: \Omega_l^m(r).
\end{equation}
The coefficients $\Omega_l^m(r)$ can be determined from simulations according to 
\begin{equation}
\Omega^m_l(r)=\frac{1}{\mathcal{V}}\sum_{n\in \delta \mathbf{r}}Y_{lm}(\mathbf{r}_n/r_n),
\label{eq:yl1m0}
\end{equation}
where the sum $\sum_{n\in \delta \mathbf{r}}$ runs over all bath particles in a spherical shell $\delta r$ 
around the particle (i.e., their distance from the probe center lies within the interval
$[r-\delta r/2: r + \delta r/2]$), and $\mathcal{V}$ is the volume of the shell, $\mathcal{V}(r)=4/3\pi((r+\delta r/2)^3-(r-\delta r/2)^3)$. Specifically, the radial average of $\rho(\mathbf{r})$ can be obtained from
\begin{equation}
\rho(r)= Y_{00}^* \: \Omega_0^0 = \frac{\sqrt{4\pi}}{\mathcal{V}}\sum_{n\in \delta \mathbf{r}}Y^0_0(\mathbf{r}_n/r_n).
\label{eq:density}
\end{equation}

\begin{figure}
  \centering
  \includegraphics[width=0.75\linewidth]{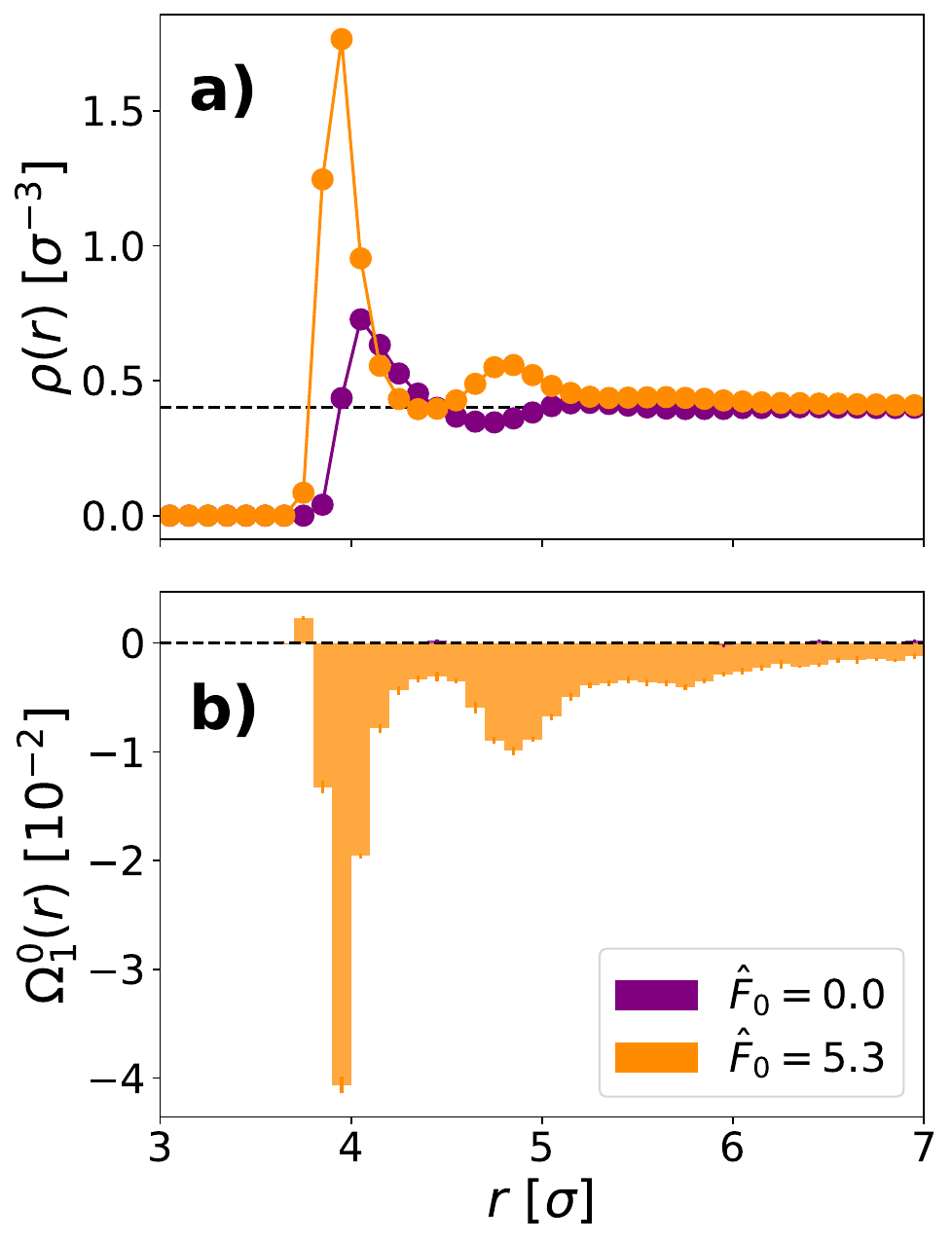}  
\caption{\textbf{a)} Radially averaged density ($\rho(r)$, see Eq.~\eqref{eq:density}) and \textbf{b)} dipole moment ($\Omega_1^0(r)$, see Eq.~\eqref{eq:yl1m0}), surrounding a free probe immersed in a bath with activities $\hat{F}_0=0.0$ (indigo) and $\hat{F}_0=5.3$ (orange). The probe has a radius $R_p=3.0\sigma$ and is immersed in a bath with global density $\rho=0.4\sigma^{-3}$. The black dashed lines mark the global density in \textbf{a})
and the zero baseline in \textbf{b}).}
\label{fig:nopot_sh}
\end{figure}

We first characterize the bath in the case that the probe is free. In comparing the radially averaged density curves for a passive and an active bath surrounding a free probe in Fig.~\ref{fig:nopot_sh}a), we see that the first peak of the density curve is higher in an active bath. Hence adding activity to the bath leads to a higher density of bath particles in the vicinity of the probe. We also see that the \js{first shell of active particles} is shifted slightly closer to the probe center in comparison with that of the passive bath. \js{Furthermore, the spacing between the first and second shells of the active bath particles are closer together.  This -- in combination with the forward shifted first shell -- leads to the coincidence of a peak in the density profile of the active bath with a trough in the density profile of the active bath at $r\approx4.9\sigma$.} \js{These closer shells indicate} that ALPs are able to move closer to the probe\js{, and each other,} than passive bath particles. We infer that this ability stems from the higher kinetic energy of ALPs due to their active force, which allows them to overcome more of the repulsive potential from WCA interactions with the probe.

In a passive bath, all higher order moments of the bath are zero. However, in an active bath, the active fluid acquires a negative dipole moment surrounding the probe, which is sustained to large values of $r$, as can be seen in Fig.~\ref{fig:nopot_sh}b). 
This negative dipole moment indicates that ALPs collect behind the probe relative to its instantaneous velocity, even at large distances from the probe. The phenomenon of active particles gathering behind an immersed passive probe has previously been seen for a probe dragged through an ABP bath in Ref.~\cite{Stark_Milos}, where it is framed as a difference between the forces behind and in front of the probe. We find no significant structure in higher order spherical harmonics for either the active or passive bath.

\begin{figure}
  \centering
  \includegraphics[width=\linewidth]{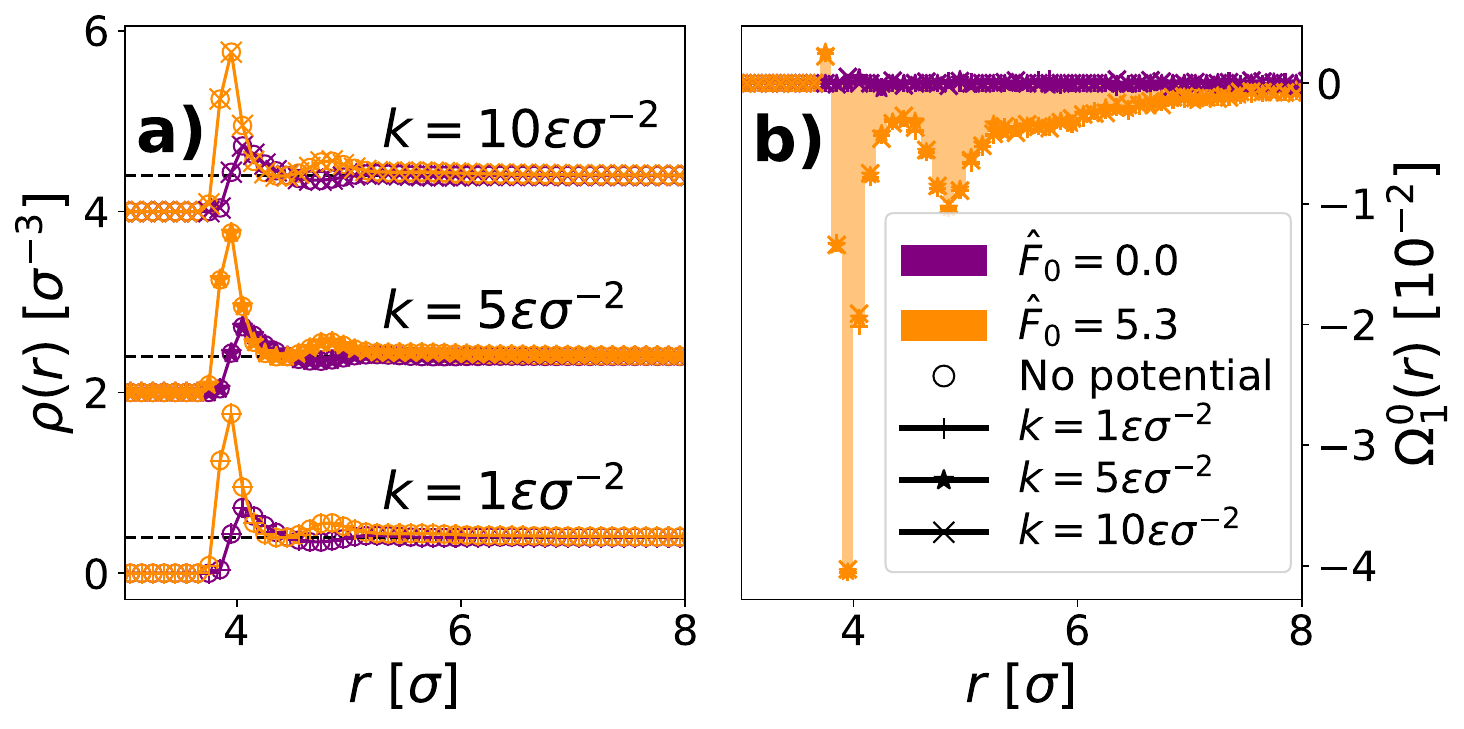}  
\caption{\textbf{a)} Radially averaged density (Eq.~\eqref{eq:density})) and \textbf{b)} dipole moment ($\Omega_1^0(r)$, see Eq.~\eqref{eq:yl1m0}), surrounding a harmonically confined probe immersed in a bath with global density $\rho=0.4\sigma^{-3}$ and activity level $\hat{F}_0=0.0$ (indigo) and $\hat{F}_0=5.3$ (orange). The probe has a radius $R_p=3.0\sigma$.  Different symbols show different harmonic trapping strengths $k=1\epsilon\sigma^{-3}$ (plus symbol), $k=5\epsilon\sigma^{-3}$ (star symbol), $k=10\epsilon\sigma^{-3}$ (x symbol). The results for a free probe are shown in \textbf{a)} by open circles and in \textbf{b)} by bars. For ease of visualization, in \textbf{a)} we add different vertical shifts to the density profiles for different trapping strengths and mark the global density with black dashed lines.  
} 
\label{fig:harm_sh}
\end{figure}

Returning to the system of a probe immersed in an active bath and subject to a harmonic constraint, we find that in spite of the additional harmonic force on the probe, the density profile of the ALP bath (as calculated from Eq.~\eqref{eq:density}) remains identical for a trapped and a free probe (see Fig.~\ref{fig:harm_sh}a)). Additionally, its dipole moment remains the same whether the probe is trapped or free (see Fig.~\ref{fig:harm_sh}b)). This is most likely due to the fact that the forces exerted by the harmonic trap are relatively small, i.e., within the linear regime. We will see in section \ref{sec:drag} that the situation changes if we subject the probe to very large forces.   

\subsection{Constant drag force}
\label{sec:drag}
\begin{figure}[b]
  \centering
  \includegraphics[width=.5\linewidth]{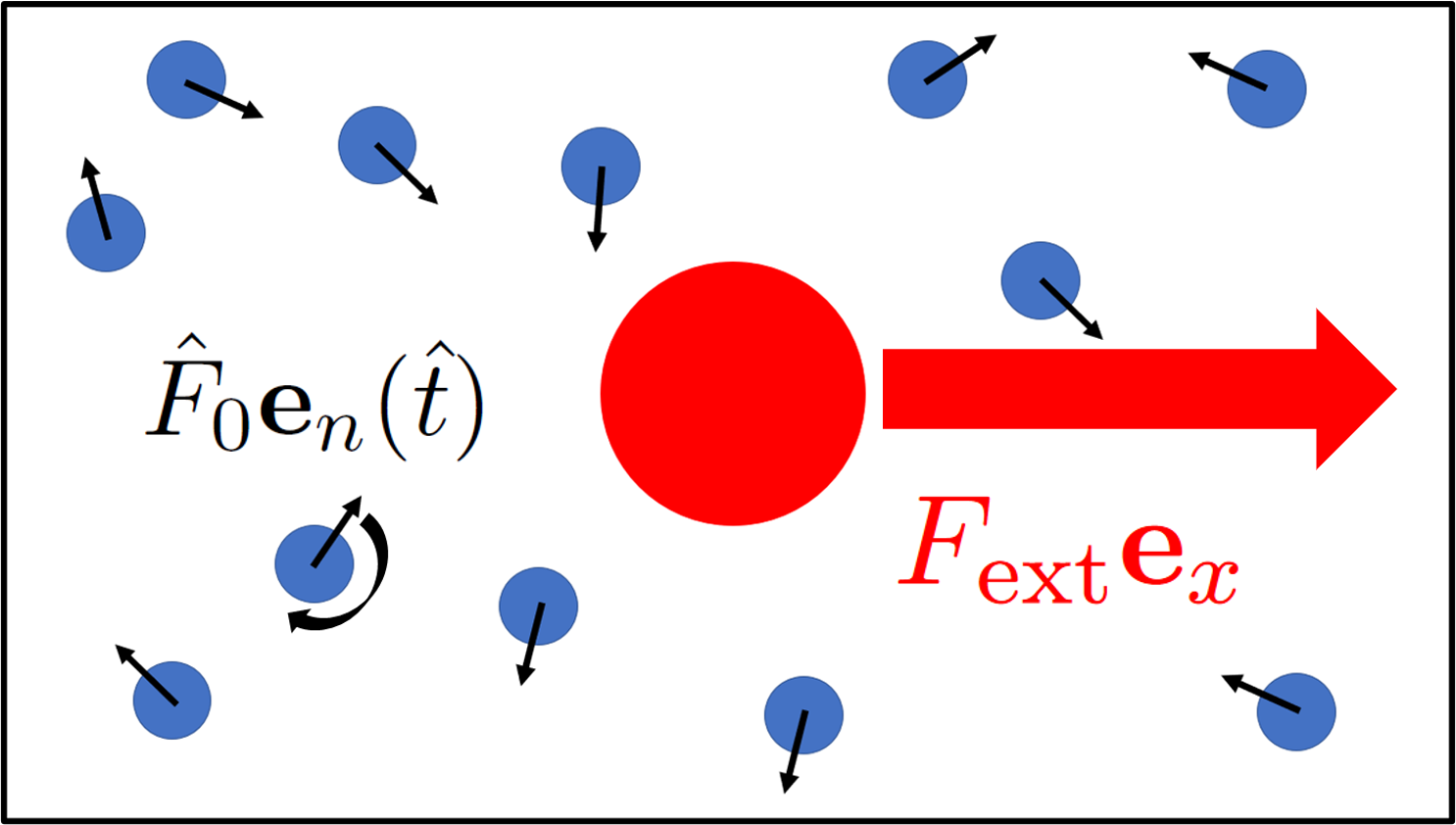}  
\caption{Cartoon of a passive probe dragged by a constant, external force through a bath of active Langevin particles.}
\label{fig:sys_pull}
\end{figure}

Next we consider a passive probe which is pulled through a bath of interacting ALPs with a constant `drag' force along the x-axis: $\mathbf{F}=F_{\mathrm{ext}}\mathbf{e}_x$ (see Fig.~\ref{fig:sys_pull}). Thus, the equation of motion for the probe is:
\begin{equation}
\label{eq:eom_coll_drag}
M \dot{\mathbf{V}}(t)=F_{\mathrm{ext}}\mathbf{e}_x-\sum_n \nabla U_\mathrm{WCA}(\mathbf{R}-\mathbf{r}_n),
\end{equation}
where $M$ is the mass of the probe, $\mathbf{V}(t)$ is its velocity, and $U_\mathrm{WCA}(\mathbf{R}-\mathbf{r}_n)$ is the WCA potential due to an ALP particle, $n$, at position $\mathbf{r}_n$. Aside from the constant external force, the immersed probe only experiences forces from interactions with the surrounding ALPs.
The ALP bath has a global density of $0.4\sigma^{-3}$. In the steady mode, the constant drag force induces a steady drift  of the probe in $x$ direction with average  velocity $\langle \mathbf{V} \rangle_{\textrm{neq}} \equiv \overline{\mathbf{V}}$. 

\gj{In our coarse-graining approach} we map this system onto a \fsnew{stationary} GLE of the form
\begin{equation}
\label{eq:gle_drift}
M \dot{\mathbf{V}}(t)=\mathbf{f}-\int^t_{\fsnew{- \infty}} \mathrm{d}s \: \mathbf{K}(t-s) \mathbf{V}(s)+\bm{\Gamma}(t),
\end{equation}
where $\mathbf{f}$ is the  effective drift force. We use a general tensorial memory kernel here as $\mathbf{K}$ may become anisotropic for large drag forces \cite{wip}.
Following Ref.\ \cite{wip}, we
split this equation into two parts:
\begin{equation}
\label{eq:gle_drift_drag}
M \dot{\bar{\mathbf{V}}}(t)= 0=\mathbf{f}-\int^t_{\fsnew{-\infty}} \mathrm{d}s \: \mathbf{K}(t-s) \overline{\mathbf{V}},
\end{equation}
\begin{equation}
\label{eq:gle_drift_frame}
M \dot{\mathbf{U}}(t)=-\int^t_{\fsnew{-\infty}} \mathrm{d}s \: \mathbf{K}(t-s) \mathbf{U}(s)+\bm{\Gamma}(t).
\end{equation}
Here, $\mathbf{V}(t)=\mathbf{U}(t)+\overline{\mathbf{V}}$ such that $\overline{\mathbf{V}}$ is the drift velocity of the probe. Eq.~\eqref{eq:gle_drift_frame} has exactly the same form as the GLE for an unconstrained probe. The only difference is that the velocity $\mathbf{U}(t)$ is now the velocity of the probe in a \emph{co-moving frame} with velocity $\overline{\mathbf{V}}$. Since the equations are of the same form, Eq.~\eqref{eq:gle_drift_frame} can also be transformed into a Volterra equation, which can then be inverted to find the memory kernel of the probe. The generalized 2FDT is fulfilled
by construction \cite{wip}. Using such a mapping, the effective drift force on the probe can then be calculated from Eq.~\ref{eq:gle_drift_drag}, as $\mathbf{f}=\int^t_{\fsnew{-\infty}} \mathrm{d}s \: \mathbf{K}(t-s)\overline{\mathbf{V}}$.

At equilibrium, the behavior of the probe at small drag velocities can be described by linear response theory. One purpose of the present study is to examine whether a similar behavior can be found in an active bath, if we renormalize the force according to the prescription found in the previous section, \ref{sec:harm}. In linear response theory, the drift velocity of the probe in response to a constant drag force $\mathbf{F}_{\textrm{ext}}$ is given by the linear relation $\langle \mathbf{V} \rangle = \mu_{\textrm{eq}} \mathbf{F}_{\textrm{ext}}$ with the mobility $\mu_{\textrm{eq}} = \big[ \int_0^\infty K_{\textrm{eq}}(s) \textrm{d}s \big]^{-1}$. Here  $K_{\textrm{eq}}$ is the equilibrium memory kernel  evaluated at $\mathbf{F}_{\textrm{ext}}=0$.

The analogous behavior in an active bath would be  $\langle \mathbf{V} \rangle = \mu_0 \mathbf{F}_{\textrm{ext}}^{\textrm{eff}}$, where $\mu_0$ is determined from the integrated
memory kernel of the free probe and $\mathbf{F}_{\textrm{ext}}^{\textrm{eff}}= \alpha \mathbf{F}_{\textrm{ext}}$ is renormalized  according to Section \ref{sec:harm}. We should note that  that linear response theory is {\em not} generally valid in our system \cite{ME}. Our goal here is to probe the existence of a linear regime where $\langle \mathbf{V} \rangle \propto \mathbf{F}_{\textrm{ext}}$, and test the relation
$\langle V \rangle/F_{\textrm{ext}} = \alpha \mu_0$ in that regime.

In Ref.~\cite{Stark_Milos}, a probe dragged through a 2D bath of ALPs has been studied for drag forces well beyond the linear regime. Here, we focus on small to intermediate forces and possible transitions between linear and nonlinear regimes.  

\subsubsection{Static mobility}
\label{sec:mobility}

In the following, we use dimensionless quantities $\hat{F}_{\textrm{ext}}= \frac{1}{\gamma} \sqrt{\frac{m}{\kbT}} F_{\textrm{ext}}$ and $\hat{\mu} = \mu \frac{1}{\gamma}$. Furthermore, to unify the notation for active and passive bath particles, we define a dimensionless renormalized force $\tilde{F}_{\textrm{ext}}$, which is given
by $\tilde{F}_{\textrm{ext}} = \hat{F}_{\textrm{ext}}$ in a passive bath
and $\tilde{F}_{\textrm{ext}} = \alpha \hat{F}_{\textrm{ext}}$ in an active bath. We will consider three values of the mobility (in dimensionless form): (i) the drag mobility $\hat{\mu}_{\textrm{Drag}}$ defined by
\begin{equation}
\label{eq:diffcoeff_drag}
\langle \hat{V}_x \rangle=\hat{\mu}_\mathrm{Drag}\Tilde{F}_\mathrm{ext},
\end{equation}
where $\langle \hat{V}_x \rangle = \hat{\overline{V}}_x$ is the dimensionless average velocity along the x-axis;

(ii) the linear GLE mobility $\hat{\mu}_0$ defined by
\begin{equation}
\label{eq:diffcoeff_k}
\hat{\mu}_0=\left[\int^\infty_0\mathrm{d}\hat{t}^{\prime}\ \hat{K}_0(\hat{t}^{\prime})\right]^{-1},
\end{equation}
where $\hat{K}_0$ is the (isotropic) dimensionless memory kernel of a free probe
(at $\Tilde{F}_\mathrm{ext}=0$);
and (iii) the tensorial GLE mobility $\hat{\bm{\mu}}_{\mathrm{K}}$ defined by
\begin{equation}
\label{eq:mobility_GLE}
\hat{\bm{\mu}}_{\mathrm{K}}=\left[\int^\infty_0\mathrm{d}\hat{t}^{\prime}\ \hat{\mathbf{K}}(\hat{t}^{\prime})\right]^{-1},
\end{equation}
where $\mathbf{K}$ is the memory kernel in the presence of the external force. To properly compare the ranges of 
the linear regime in the passive and active bath case, we introduce a dimensionless Peclet number for the probe: $\mathrm{Pe}=\langle \hat{V}_x \rangle/\hat{V}_\mathrm{diff}$, where the diffusion velocity scale $\hat{V}_\mathrm{diff}$
is given by $\hat{V}_\mathrm{diff}=D_\mathrm{eff} \sqrt{{m}/{\kbT}}/R_p$ and depends on the probe radius $R_p=3.0\sigma$ and the effective diffusion coefficient of an isolated ALP,  $D_\mathrm{eff}^{ALP}=\kbTe^{\textrm{ALP}}/\gamma$ with $\kbTe^{\textrm{ALP}}=\kbT [1 + \hf^2/(3+6 \hdr)]$ as calculated in Ref.~\cite{ME}. In terms of $\Tilde{F}_\mathrm{ext}$, the Peclet number is given by $\mathrm{Pe}=\hat{\mu}_\mathrm{Drag}\Tilde{F}_\mathrm{ext}/\hat{V}_\mathrm{diff}$ (see Appendix~\ref{sec:app_pefext}).


In Fig.~\ref{fig:mobility}a), we graph $\langle \hat{V}_x \rangle$ as a function of $\Tilde{F}_\mathrm{ext}$ to determine the drag mobility $\hat{\mu}_\mathrm{Drag}$ within the linear regime using Eq.~\eqref{eq:diffcoeff_drag}. \js{We note that, for low values of $F_\mathrm{ext}$ in an active bath, the stochastic motion of the active bath particles dominates the applied external force. Hence, the deviation of $\langle \hat{V}_x \rangle$ is significantly larger for low values of $F_\mathrm{ext}$, resulting in larger errors in our data.}

We extract a mobility $\hat{\mu}_\mathrm{Drag}=0.02057 \pm 0.00006$ for a probe immersed in a passive bath ($\hf=0.0$) with a density $\rho=0.4\sigma^{-3}$. For a probe in an active bath ($\hf=5.3$) of density $\rho=0.4\sigma^{-3}$, we extract a mobility of $\hat{\mu}_\mathrm{Drag}=0.154 \pm 0.006$. Comparing these values, we see that increasing the activity of the bath also increases the mobility of the probe particle.

On the other hand, using the memory kernels of a free probe and Eq.\ (\ref{eq:diffcoeff_k}), we obtain the
linear GLE mobility $\hat{\mu}_0=0.0226 \pm 0.0003$ for a probe in a passive bath of density $\rho=0.4\sigma^{-3}$ and $\hat{\mu}_0=0.147 \pm 0.002$ for a probe in an active bath of density $\rho=0.4\sigma^{-3}$ with $\hf=5.3$. These values for the mobility agree well with those calculated from Eq.~\eqref{eq:diffcoeff_drag}. The slight discrepancy between values is most likely due to the reconstruction of the memory, which becomes less accurate at longer times. We emphasize that renormalizing the drift force in Eq. (\ref{eq:diffcoeff_drag}) is essential for reaching this agreement. 
\gjc{The connection between effective temperature and effective drift reminds me of \cite{Szamel2021_efftemp}.}

\begin{figure*}[t]
  \centering
  \includegraphics[width=1.\linewidth]{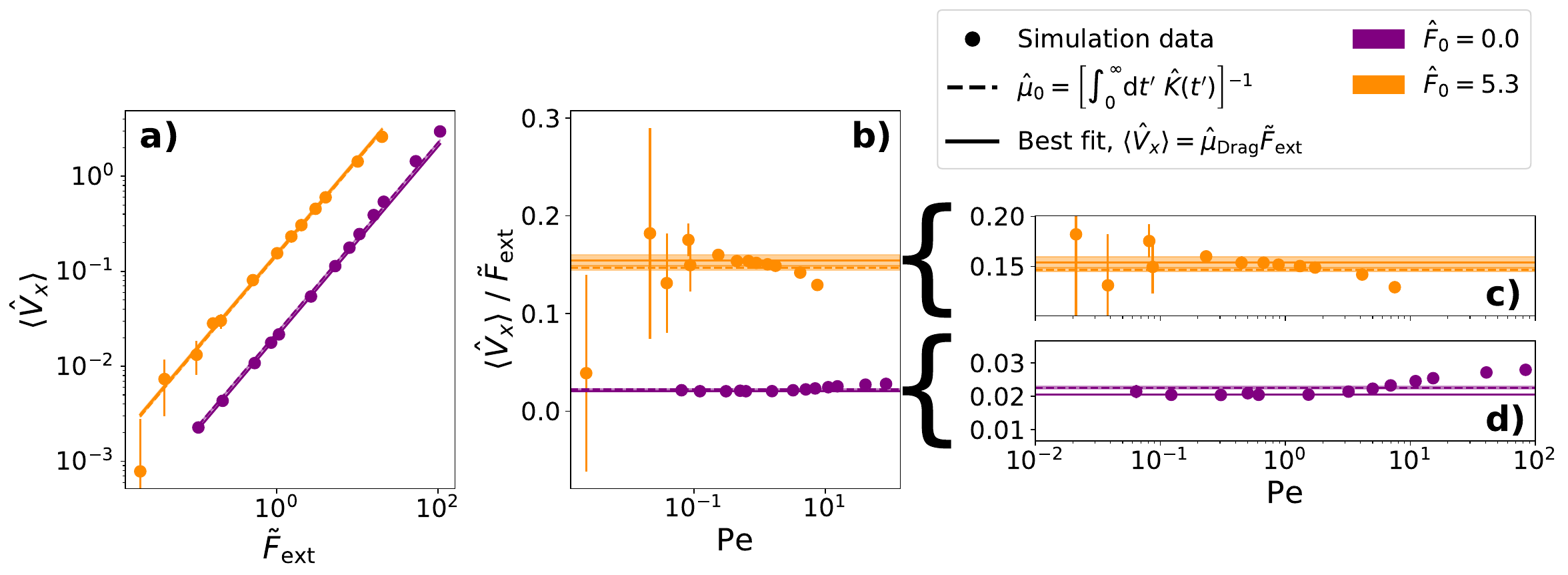}
\caption{Drag mobility of the probe as a function Peclet number for probes in an active (orange) and passive (indigo) bath. The circles show the simulation data. \textbf{a)} Average velocity along the x-axis, $\langle \hat{V}_x \rangle$, as a function of the renormalized external drag force, $\Tilde{F}_\mathrm{ext}$. The solid lines show the best linear fit for these data, the slope of which is $\hat{\mu}_\mathrm{Drag}$ in the linear regime. \textbf{b)} Force dependent mobility as a function of Peclet number. \textbf{c)}, \textbf{d)} Zoom in on the mobility as a function of Peclet number in the \textbf{c)} active case and \textbf{d)} passive case (in \textbf{c)}, the first data point is not shown).  The dashed lines mark the value of $\hat{\mu}_{\mathrm{K}}$. Light shading indicates error bars. Each bath has average density $\rho=0.4\sigma^{-3}$ and the probe has a radius $R_p=3.0\sigma$.}
\label{fig:mobility}
\end{figure*}

For high forces, the drag mobility is no longer a fixed value. In Fig.~\ref{fig:mobility}b), we show the force dependent drag mobility ($\hat{\mu}_\mathrm{Drag}(\Tilde{F}_\mathrm{ext})=\langle \hat{V}_x \rangle/\Tilde{F}_\mathrm{ext}$) as a function of the Peclet number. We expect $\hat{\mu}_\mathrm{Drag}(\Tilde{F}_\mathrm{ext})$ to be constant within the linear response regime for a probe immersed in a passive bath. The figure shows that this is indeed the case for Peclet number up to $\mathrm{Pe}\sim1$.
For probes immersed in an active bath, we observe a very similar behavior, i.e.,  a crossover from roughly constant drag mobility to force dependent drag mobility around $\mathrm{Pe}\sim1$. Identifying this crossover is more \js{difficult} in this case because the measured values for $\hat{\mu}_{\textrm{Drag}}$ fluctuate strongly for low values of Pe.
These fluctuations reflect the fact that the activity of the bath induces larger variations in the probe velocity.

We should note that, due to the enhanced diffusion of the active bath particles in comparison to the passive bath particles, equivalent Peclet numbers for a probe in an active and a passive bath correspond to different external drag forces (see also Appendix~\ref{sec:app_pefext}). Much larger drag forces are required to achieve the same Peclet number for an active bath in comparison with a passive bath.

\label{sec:beyond_lin}
Figs.~\ref{fig:mobility}c) and d) show that the behavior beyond the linear response regime is qualitatively different for a probe immersed in an active vs. a passive bath.  A probe dragged through a passive bath exhibits thinning behavior (increased mobility), whereas a probe dragged through a bath of ALPs exhibits thickening behavior (reduced mobility). 
Due to these opposite behaviors, the values of $\hat{\mu}_\mathrm{Drag}(\Tilde{F}_\mathrm{ext})$ for a passive and an active bath at the same Peclet number approach each other beyond the linear regime. \fs{We will discuss and analyze this further at the end of Section \ref{sec:gle_comove}}. 


\subsubsection{Velocity autocorrelation function and memory kernel}
\label{sec:drag_mk}
Having looked at the active microrheological properties of the active bath, we turn to the dynamic properties of the dragged probe: namely, the VACF and the memory kernel. We map the dynamics of our system onto Eq.~\eqref{eq:gle_drift}. We examine the VACF in the co-moving frame because, for a probe dragged through a passive bath, it should be roughly identical for all drag forces within the linear response regime~\cite{wip}. We would like to evaluate the universality of the co-moving VACF for a probe dragged through an active bath as well.

\begin{figure}
  \centering
  \includegraphics[width=.887\linewidth]{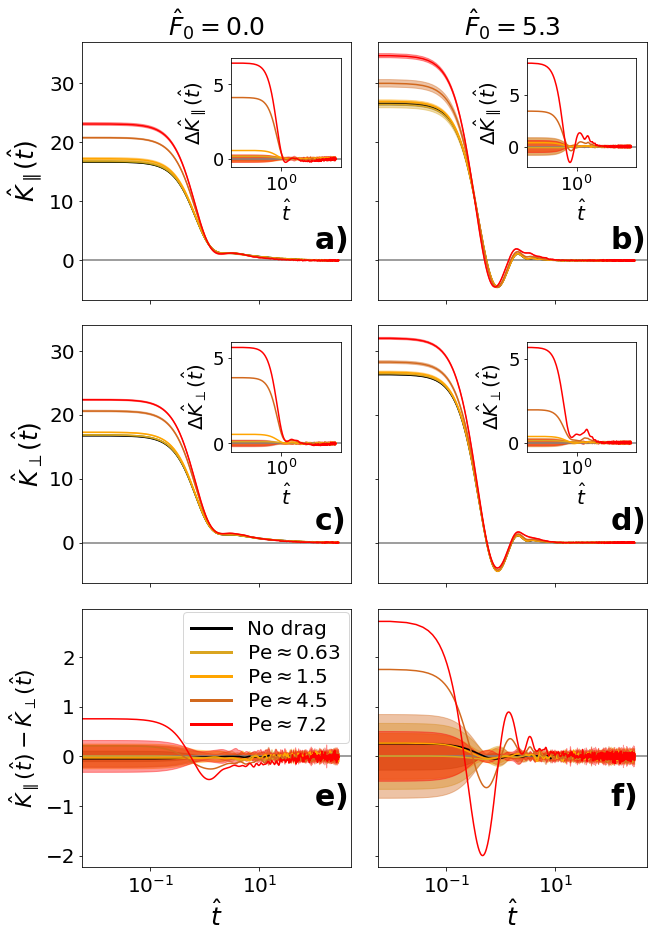}  
\caption{Memory kernel of the dragged, immersed probe. Each bath has global density $\rho=0.4\sigma^{-3}$ and the probe has a radius $R_p=3.0\sigma$. \textbf{a},\textbf{b)} The top row shows the component of the memory kernel which is parallel to the drag force, whereas \textbf{c},\textbf{d)} the middle row shows the perpendicular component. \textbf{e},\textbf{f)} The bottom row shows the difference between parallel and perpendicular components of the memory kernel. \textbf{a},\textbf{c},\textbf{e)} The left column is for a passive bath, whereas \textbf{b},\textbf{d},\textbf{f)} the right column is for an active bath ($\hat{F}_0=5.3$). The black curves in plots \textbf{a}-\textbf{d)} show the memory kernel for a probe which is not dragged and insets show the difference between the memory kernel of a free and a dragged probe. Light shading indicates error bars. 
}
\label{fig:drag_mk}
\end{figure}

As we would expect, within the linear regime ($\mathrm{Pe}<1$), the VACF and memory kernel are isotropic and independent of $\Tilde{F}_\mathrm{ext}$ for a probe dragged through a passive bath. \js{We note that, in Figs.~\ref{fig:drag_mk}a) and b), the memory kernel for a probe dragged with $\mathrm{Pe}\approx1.5$ through a passive bath already deviates slightly from the memory kernel of the undragged probe. This suggests that the transition from the linear to the nonlinear regime in a passive bath is a relatively sharp transition at $\mathrm{Pe} = 1$. Because these deviations are primarily visible in the memory kernel as opposed to the VACF, this suggests that the transition to the nonlinear regime is primarily due to the stochastic force distribution. In fact, for a probe dragged through a passive bath at $\mathrm{Pe}\approx1.5$, we can already see a slight asymmetry in the probe's stochastic force distribution (see Appendix~\ref{sec:app_sfpe15}). This asymmetry is a signature of the nonlinear regime, as will be discussed in the next section,~\ref{fig:drag_sf}.}

\begin{figure}
  \centering
  \includegraphics[width=1.\linewidth]{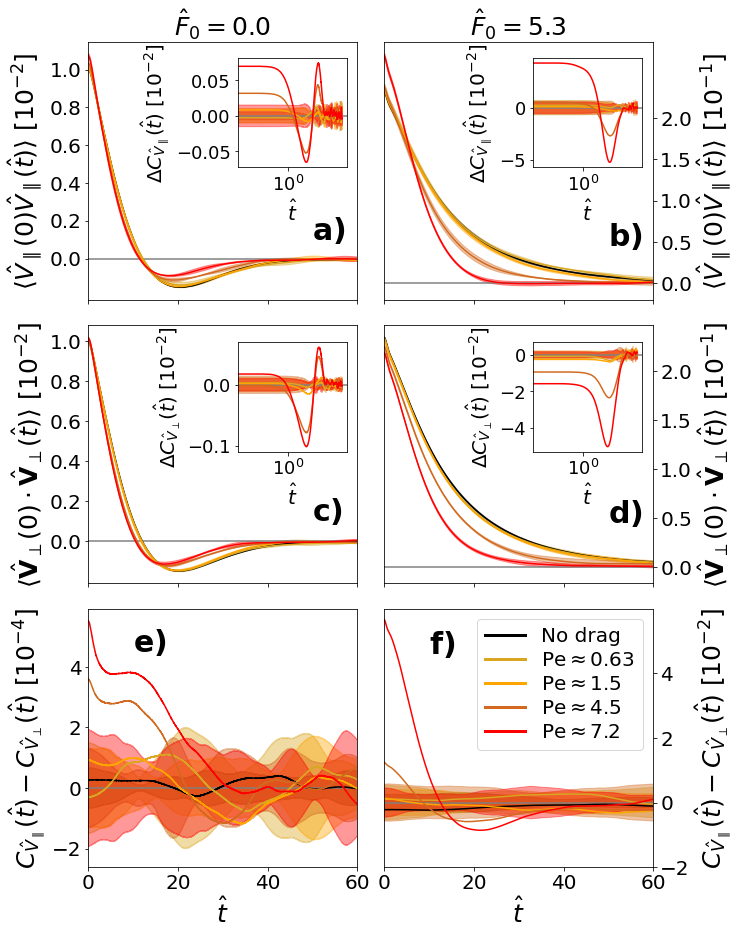}  
\caption{Velocity autocorrelation function (VACF) of the dragged, immersed probe. Each bath has global density $\rho=0.4\sigma^{-3}$ and the probe has a radius $R_p=3.0\sigma$. \textbf{a},\textbf{b)} The top row shows the component of the VACF which is parallel to the drag force, whereas \textbf{c},\textbf{d)} the middle row shows the perpendicular component. \textbf{e},\textbf{f)} The bottom row shows the difference between parallel and perpendicular components of the memory kernel. \textbf{a},\textbf{c},\textbf{e)} The left column is for a passive bath, whereas \textbf{b},\textbf{d},\textbf{f)} the right column is for an active bath ($\hat{F}_0=5.3$). The black curves in plots \textbf{a}-\textbf{d)} show the VACF for a 
free probe and insets show the difference between the VACF of a free and a dragged probe. Light shading indicates error bars. 
}
\label{fig:drag_vv}
\end{figure}

\js{For the active bath, we also see that the VACF and memory kernel are isotropic and independent of $\Tilde{F}_\mathrm{ext}$ within the linear regime. Furthermore, in this case, the linear regime appears to hold for Peclet numbers of up to at least $\mathrm{Pe}\approx1.5$ (Figs.~\ref{fig:drag_vv} and \ref{fig:drag_mk}), suggesting that the tranistion from the linear to nonlinear regime is less sharp in an active bath.}

Beyond the linear regime, however, the VACF and memory kernel are neither isotropic (see Figs.~\ref{fig:drag_vv}e,f) and~\ref{fig:drag_mk}e,f)) nor independent of $\Tilde{F}_\mathrm{ext}$ for both passive and active systems. In both cases, a large drag force causes the VACF to decay more rapidly in both its parallel and perpendicular components. The memory kernel is primarily affected at short times, where it is greater than the memory kernel of a free probe. This is, again, true for both a probe dragged through a passive and an active bath.
However, the other aspects of the VACF/memory kernel form differ for passive and active baths. In a passive bath, we see that the dragged probe acquires an increased local kinetic temperature in both its parallel and perpendicular components. In an active bath, on the other hand, the dragged probe only acquires an increased local kinetic temperature in its parallel component. In its perpendicular component, the probe dragged through an active bath actually experiences a decrease in local kinetic temperature.

As a consequence of the changes in the memory kernel, the GLE mobility becomes anisotropic at high Peclet
numbers, and both its parallel and perpendicular components increase (data not shown). In the linear regime, 
however, the tensorial and linear GLE mobility are identical\js{:} 
$\hat{\bm{\mu}}_\textrm{K} = \bm{1} \hat{\mu}_0$ for $\mathrm{Pe} < 1$, where $\mathbf{1}$ is the unit tensor.

\subsubsection{Stochastic force distribution}
\label{sec:drag_sf}

\begin{figure}
  \centering
  \includegraphics[width=\linewidth]{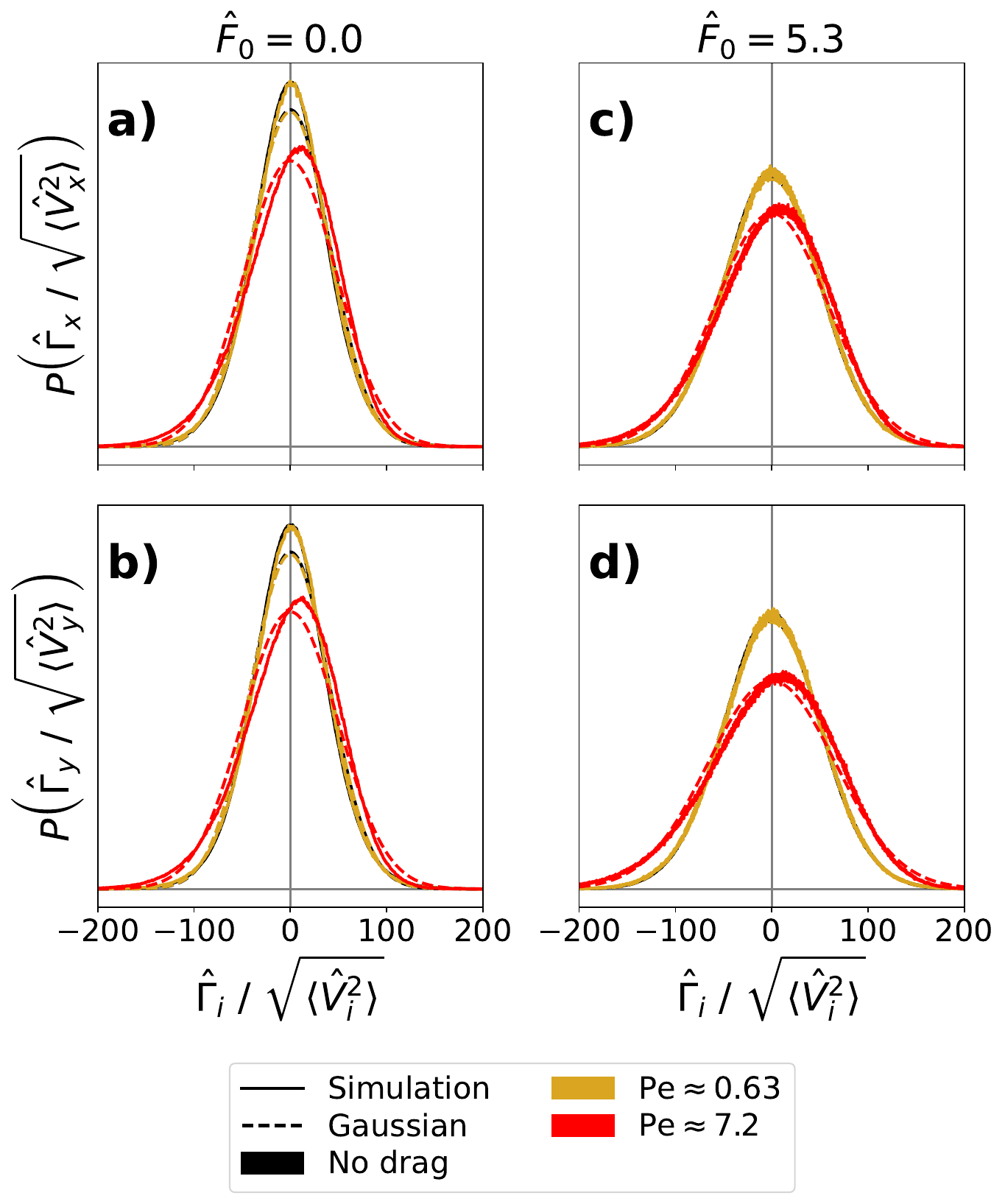}  
\caption{Stochastic force distribution of the dragged, immersed probe in a bath of density $\rho=0.4\sigma^{-3}$. \textbf{a},\textbf{b)} The top row shows the component of the stochastic force which is parallel to the drag force, whereas \textbf{c},\textbf{d)} the bottom row shows one of the perpendicular components (the other perpendicular component is identical). \textbf{a},\textbf{c)} The left column is for a passive bath, whereas \textbf{b},\textbf{d)} the right column is for an active bath ($\hat{F}_0=5.3$). The solid lines show simulation data, whereas the dotted lines show zero-centered Gaussian distributions with the same standard deviation. 
}
\label{fig:drag_sf}
\end{figure}

Once we have determined the memory kernel of the probe, we can determine the stochastic force through a trivial re-ordering of Eq.~\eqref{eq:gle_drift_frame}~\cite{Shin_2010}:
\begin{equation}
\label{eq:stoch_drift}
\bm{\Gamma}(t)=\mathbf{F}(t) +\int^t_{\fsnew{-\infty}} \mathrm{d}s \: \mathbf{K}(t-s) \mathbf{U}(s).
\end{equation}
where $\mathbf{F}(t)$ is the instantaneous force acting on the particle, which is calculated during the simulated trajectory. The probability distribution of the stochastic force is shown in Fig.~\ref{fig:drag_sf}. In both the case of a passive and an active bath, for low Peclet numbers (within the linear regime), the stochastic force distribution calculated from simulation data (solid lines) approximately matches a zero-centered Gaussian with the same standard deviation (dashed lines). The only deviations from the zero-centered Gaussian are, again, an enhanced peak in the distribution, particularly in the case of the passive bath, which we infer results from the low density of the bath. 

However, at higher drag forces, the distribution calculated from simulation data develops an asymmetry in both components parallel and perpendicular to the drag force. This asymmetry has already been seen for a passive bath in Ref.~\cite{wip}. Our results here show that it also occurs for an active bath, although the magnitude of the asymmetry appears to be lower in the active case. We infer that the lower asymmetry in the active case stems from the already enhanced fluctuations due to the bath activity. 

The asymmetry takes the form of a long tail in the negative forces of the distribution. This long tail occurs because, outside of the linear regime, bath particles in the vicinity of the probe do not necessarily travel at the same relative velocity as the probe~\cite{squires}. This leads some bath particles to `crash,' with relatively high forces, into the probe, promoting the long tail. Because the average of the stochastic force distribution remains zero, the distribution also exhibits a slightly enhanced probability for positive forces. The distribution outside the linear regime can be described by a split normal Gaussian~\cite{wip}. 

\subsubsection{Characterization of the active bath}
\label{sec:sh_mk}

\begin{figure}
  \centering
  \includegraphics[width=.75\linewidth]{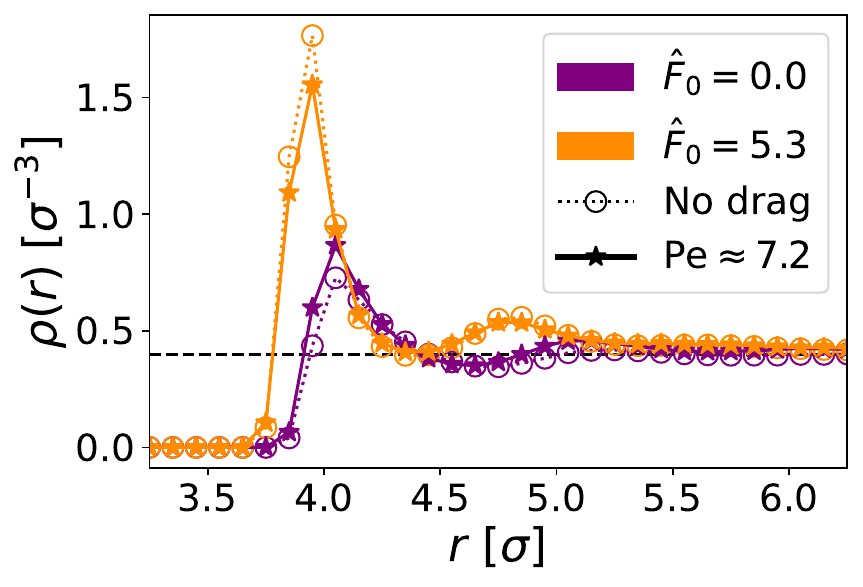}  
\caption{Radially averaged density, see Eq.~\eqref{eq:density}) of the immersed probe in baths with activities $\hat{F}_0=0.0$ (indigo) and $\hat{F}_0=5.3$ (orange). Open circles with dotted lines show $\rho(r)$ for a probe which is not dragged, whereas solid lines with stars show $\rho(r)$ for a probe dragged with $\mathrm{Pe}\approx7.2$. The probe has a radius $R_p=3.0\sigma$. The black dashed line marks the global density, $\rho=0.4\sigma^{-3}$.}
\label{fig:drag_dens}
\end{figure}

We hypothesize that the behavior of the probe mobility beyond the linear response regime results primarily from the different number of contacts experienced by a dragged probe. Fig.~\ref{fig:drag_dens} shows that  both for a passive and an active bath, the density profiles (calculated from Eq.~\eqref{eq:density})  for a free vs. a dragged probe particle differ from each other. However, the effect of the drag force on the density profile is opposite in the two cases: Whereas a probe dragged through a passive bath has a higher number of contacts in its immediate vicinity, a probe dragged through an active bath has a lower number of contacts. 

\begin{figure}
  \centering
  \includegraphics[width=1.\linewidth]{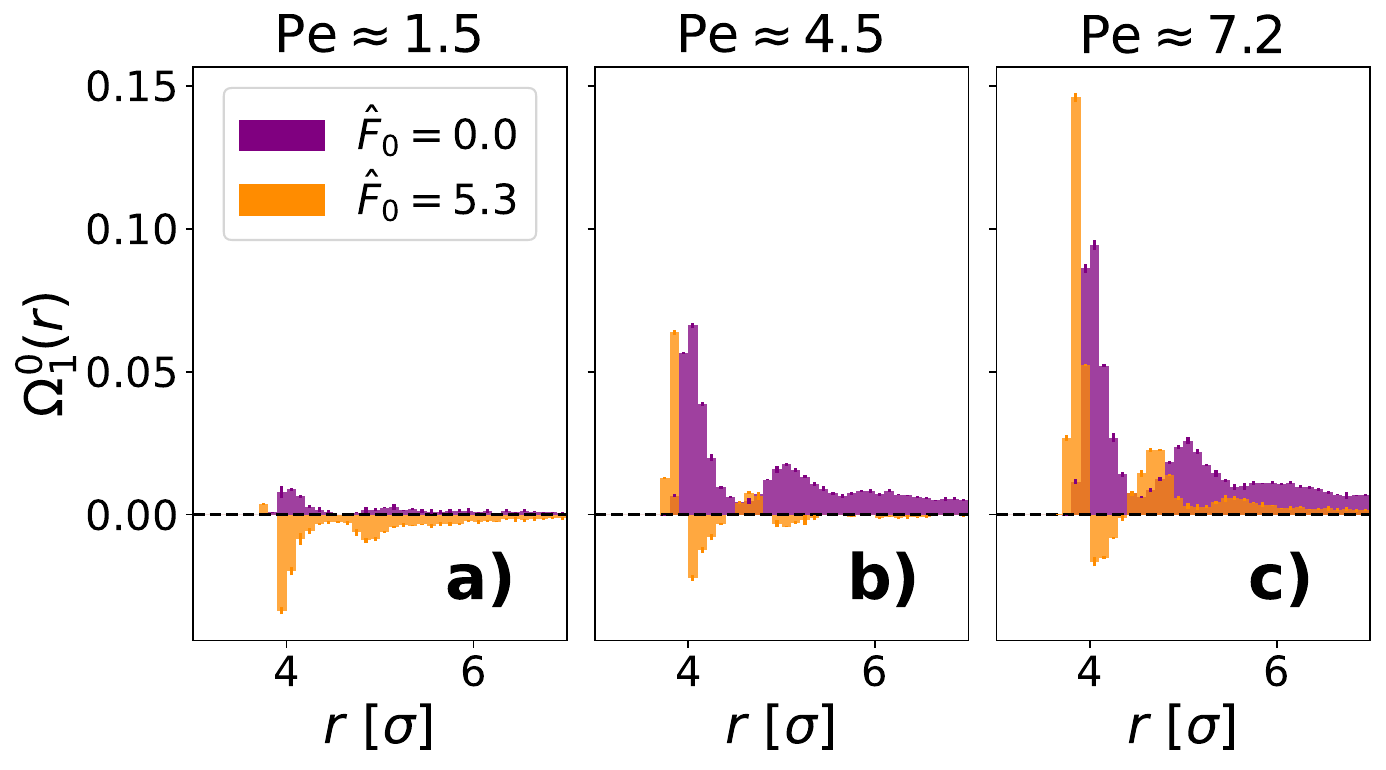}  
\caption{Dipole moment ($\Omega_1^0(r)$, see Eq.~\eqref{eq:yl1m0}) of the bath as a function of distance from the center of the immersed probe ($r$) for a probe dragged with a Peclet number \textbf{a)} $\mathrm{Pe}\approx1.5$, \textbf{b)} $\mathrm{Pe}\approx4.5$, and \textbf{c)} $\mathrm{Pe}\approx7.2$ for both a passive ($\hat{F}_0=0.0$, indigo) and an active ($\hat{F}_0=5.3$, orange) bath. Each bath has average density $\rho=0.4\sigma^{-3}$ and the probe has a radius $R_p=3.0\sigma$. The black dashed line marks the zero baseline.
}
\label{fig:drag_ylm}
\end{figure}

It has previously been shown that a probe dragged with sufficient force through a passive bath experiences an increased density of particles in front of it~\cite{wip,thin_thick}. By examining the dipole moment in the vicinity of the probe (see Fig.~\ref{fig:drag_ylm}, calculated from Eq.~\eqref{eq:yl1m0}), we verify this buildup of particles in front of a probe dragged through a passive bath with $\mathrm{Pe}>1$. 
A similar frontal buildup is exhibited by a probe dragged through an active bath with $\mathrm{Pe}\approx7.2$. However, for  lower Peclet numbers ($\mathrm{Pe}\approx1.5$ and $\mathrm{Pe}\approx4.5$), it is no longer observed. In spite of the drag force, we still see an accumulation of ALPs \emph{behind} the probe for $\mathrm{Pe}\lesssim1.5$.

We reason this qualitative difference in the bath behavior at the probe interface following the arguments of Ref.~\cite{Stark_Milos}. 
For small Peclet numbers, the speed of the ALPs is greater than that of the dragged probe. Thus, the ALPs are able to accumulate behind the probe and push, as was the case for a probe without an external drag force. Consequently, we still see a negative dipole moment in the active bath for $\mathrm{Pe}\approx1.5$ in Fig.~\ref{fig:drag_ylm}a). For intermediate Peclet numbers, the dragged probe and the ALPs move at similar speeds. Therefore, the dipole moment of the active bath approaches zero, which we see happening for $\mathrm{Pe}\approx4.5$ in Fig.~\ref{fig:drag_ylm}b). Although for this value of $\mathrm{Pe}$ the dipole moment is not entirely zero, we can see that its magnitude and range are smaller. For large Peclet numbers, ALPs can no longer catch up to the probe and, consequently, the active bath acquires a positive dipole moment in the vicinity of the probe, as seen for $\mathrm{Pe}\approx7.2$ in Fig.~\ref{fig:drag_ylm}c).

\subsubsection{Effective drift force at high drag forces}
\label{sec:gle_comove}
To further characterize the transition from the linear to the nonlinear regime, we now evaluate the dimensionless effective drift force in our coarse-grained GLE,
 \begin{equation}
\label{eq:eff_drift_force}
\hat{f}_x= \langle \hat{V}_x \rangle / \mu_{\textrm{K},\parallel}
 = \langle \hat{V}_x \rangle \:\int^{\infty}_0 \mathrm{d}t^{\prime} \: \hat{K}_{\parallel}(t^{\prime}).
\end{equation}
Here, $\hat{K}_{\parallel}(t)$ is the parallel component of the dragged probe's memory kernel shown in Figs.~\ref{fig:drag_mk}a) and b) and we have exploited the fact that $\hat{\mathbf{K}}$ and hence
$\hat{\bm{\mu}}$ are diagonal. 
In the linear regime, we have already  established $\hat{\mu}_{\textrm{Drag}} 
 = \hat{\mu}_{\textrm{K},\parallel} = \hat{\mu}_0$ in Secs \ref{sec:mobility} and \ref{sec:drag_mk}, hence $\hat{f}_x = \tilde{F}_{\textrm{ext}}$. Our goal in this section is to test this relation in the nonlinear regime. To this end, we plot ${\cal C} \hat{f}_x$ versus $\textrm{Pe}={\cal C} \tilde{F}_{\textrm{ext}}$ in Fig.~\ref{fig:drag_norm}.
 Here we rescale by $\mathcal{C}=\hat{\mu}_\mathrm{Drag}/\hat{V}_\mathrm{diff}$ (see Appendix~\ref{sec:app_pefext})
 in order to be able to compare the nonlinear effects in the passive and active bath case.

The figure shows that the relation $\hat{f}_x = \tilde{F}_{\textrm{ext}}$ holds in the 
linear regime as expected, but breaks down at high Peclet numbers. In the case of the 
a passive bath, $\hat{f}_x$ is smaller than $\tilde{F}_{\textrm{ext}} = \hat{F}_{\textrm{ext}}$ at large drag forces. In the case of an active bath, $\hat{f}_x$ is larger than $\tilde{F}_{\textrm{ext}} = \alpha \hat{F}_{\textrm{ext}}$. As a result, the values
of $\hat{f}_x$ in the active and passive bath at given physical drag force $\hat{F}_{\textrm{ext}}$ approach each other. This is shown in the inset of 
Fig.~\ref{fig:drag_norm}, where the ratio $\hat{f}_x/\hat{F}_{\textrm{ext}}$ is plotted
as a function of the high Peclet number.

\begin{figure}
  \centering
  \includegraphics[width=.95\linewidth]{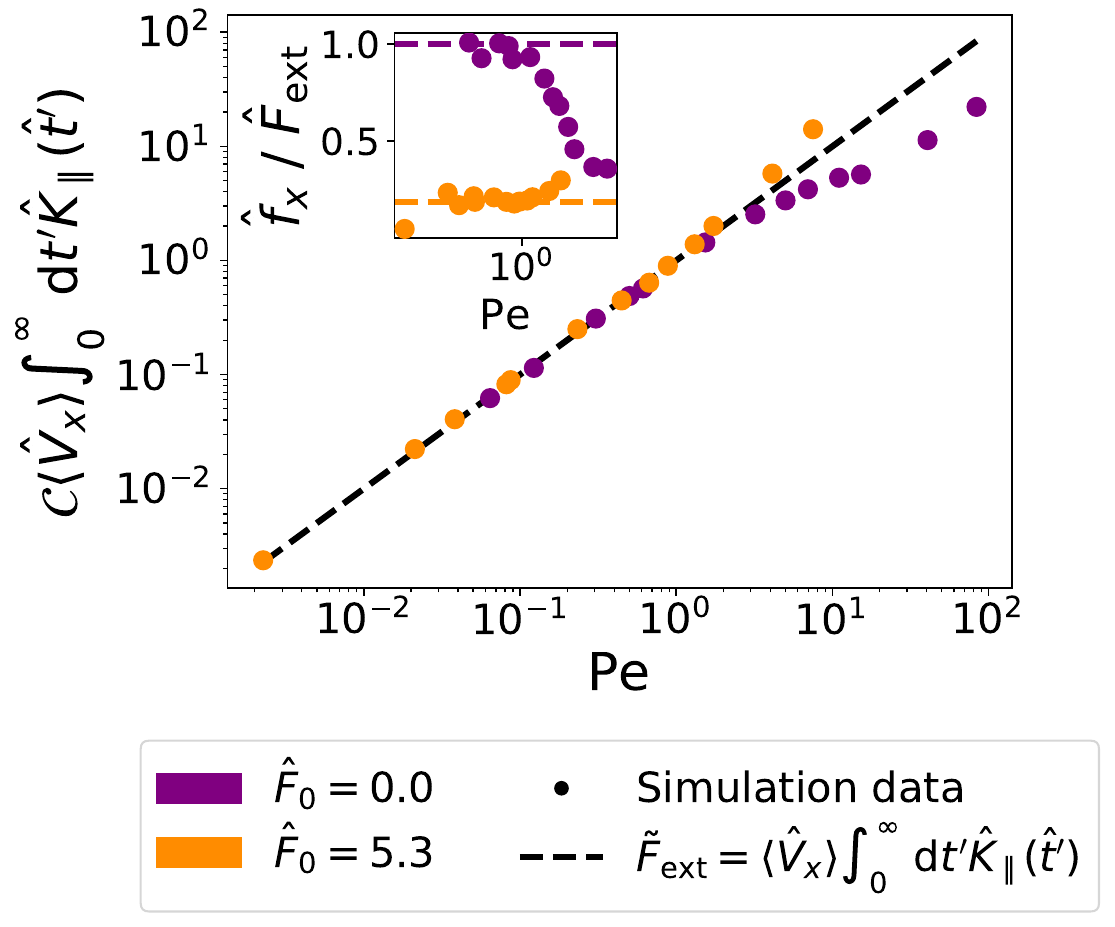}  
\caption{Effective drift force ($\hat{f}_x$) as a function of the Peclet number, rescaled
by $\mathcal{C}=\mu_\mathrm{Drag}/\hat{V}_\mathrm{diff}$ (see text for explanation)}. $\mathrm{Pe}=\mathcal{C}\Tilde{F}_{\mathrm{ext}}$ (see Appendix~\ref{sec:app_pefext}). The dashed, black line shows expected behavior $\hat{f}_x= \Tilde{F}_\mathrm{ext}$ in the
linear regime.  The inset shows $f_x/F_{\mathrm{ext}}$ as a function of the Peclet number, where the dashed lines mark $f_x/F_{\mathrm{ext}}=1$ (indigo) and $f_x/F_{\mathrm{ext}}=\alpha_{0.4}$ (orange). Each bath has average density $\rho=0.4\sigma^{-3}$ and the probe has a radius $R_p=3.0\sigma$.
\label{fig:drag_norm}
\end{figure}


We conclude that the change of the drag mobility $\hat{\mu}_{\textrm{Drag}}=\langle \hat{V}_x \rangle / F_{\textrm{ext}}$ in the nonlinear regime, which has been 
reported in Fig.~\ref{fig:mobility}, can be decomposed into two factors: First the GLE
mobility $\hat{\mu}_K$ decreases with increasing Peclet number, both in a passive and
in an active bath (cf. Fig.\ \ref{fig:drag_mk}). Second, the force renormalization factor
$\hat{f}_x/\hat{F}_{\textrm{ext}}$ increases in an active bath, and decreases in a passive 
bath. In the nonlinear regime, $F_{\textrm{ext}}$ thus must be renormalized even in the
passive system. In a passive bath, both factors result in thinning behavior. In an active bath, they compete with each other, but the force renormalization effect dominates; hence one effectively observes thickening. The latter can be associated with the redistribution of bath particles around the probe, from back end to front end, shown in 
Fig.\ \ref{fig:drag_ylm}. 

\js{In the literature, both `thinning' (increased mobility)~\cite{thinning1,thinning2,Puertas_2014,Harrer_2012,thin_thick,Jayaram_2023,Burkholder_Brady2,Stark_Milos} and `thickening' (decreased mobility)~\cite{wip,Puertas_2014,thin_thick,thickening1,thickening2} behavior have been observed for different systems beyond the linear regime. In general, systems which are not dominated by hydrodynamics exhibit thinning behavior~\cite{Puertas_2014}. This aligns with our finding that a probe dragged through a passive bath exhibits thinning behavior. However, in spite of the fact that our system neglects hydrodynamics, we find that a probe dragged through a bath of ALPs exhibits thickening behavior. This thickening behavior is in contrast to what has been previously observed in an ABP bath, where particle inertia is also neglected~\cite{Jayaram_2023,Burkholder_Brady2,Stark_Milos}. Thickening has, however, been observed in microrheological studies of granular active systems, where inertia plays an important role~\cite{thickening2,Fiege_2012_granular}.}


\section{Discussion and outlook}
\label{sec:conc_ext}

In sum, we have analyzed the dynamics of a probe particle immersed in a bath of active Langevin particles in the presence of external forces by molecular dynamics simulations. We found that it is possible to map this system onto a GLE that satisfies a 2FDT. However, in order to achieve full consistency, we must  
\begin{enumerate}
    \item[(i)] rescale the temperature such that $\kbTe = M \langle \mathbf{V}^2 \rangle/d$, with $d$ being the dimensionality of the system.
    More generally, we must replace the equilibrium 2FDT by a generalized 2FDT, 
    Eq.\ (\ref{eq:2fdt_gen}). 
    \item[(ii)] renormalize the external force, $\mathbf{F}_{\textrm{ext}}^{\textrm{eff}}= \alpha \mathbf{F}_{\textrm{ext}}$. It must be chosen such that the  distribution of probe positions in a confining renormalized potential $U^{\textbf{eff}}(\mathbf{R})$ satisfies a generalized equipartition theorem, Eq.~(\ref{eq:equipartition}). We have identified a linear regime of small to moderate forces where the 
    renormalization factor $\alpha$ does not depend on the amplitude of the force. 
\end{enumerate}
If these two conditions are fulfilled, the positions of the probe in a confining potential are Boltzmann distributed according to a Boltzmann-like distribution, 
$P(\mathbf{R}) \propto \exp(- U^ {\textrm{eff}}/\kbTe)$, the mobility of the particle can be determined consistently as semi-infinite integral over the memory kernel, and the 2FDT ensures that the mobility $\mu$ and long-time diffusion constant $D$ of a free particle are connected by a generalized Einstein relation, $D = \mu \: \kbTe$ (see Eq.~(\ref{eq:einstein}) in the appendix). 

The ability to map this non-equilibrium system onto a GLE that satisfies the generalized
2FDT enables the use of many recently developed coarse-grained methods which rely on its fulfillment~\cite{Li,Wang,GLD}. 

We emphasize that both temperature and force renormalization are necessary to achieve full consistency. This is because the effective temperature and the force renormalization depend on each other via the generalized equipartition theorem. On less formal grounds, we can also argue as follows:  If we only rescale the temperature, as is commonly done, the probe particle will have the correct velocity distribution in the GLE model, but the position distribution will deviate from the true distribution. If we only renormalize the potential, the probe will not have the correct velocity distribution.

It is instructive to discuss our findings in relation to another popular type of coarse-grained model for particles immersed in an active fluid, i.e., the overdamped Markovian Langevin equation (OLE) \cite{Jayaram_2023}. An OLE for such a system should reproduce three target quantities: (i) The diffusion constant of a free probe, (ii) the drift velocity in the presence  of a slowly varying drag force, (iii) the position distribution in the presence of a confining potential. Since the OLE can be derived from a GLE in certain limits, we can
use our results to propose a suitable OLE:
\begin{equation}
 \dot{\mathbf{R}} = \mu_K \alpha \mathbf{F}_{\textrm{ext}}  + \bm{\Xi}(t)
                  =: \tilde{\mu} \mathbf{F}_{\textrm{ext}} + \bm{\Xi}(t).
\end{equation}
Here $\bm{\Xi}(t)$ is a Gaussian distributed white noise satisfying  
$\langle \bm{\Xi}(t) \bm{\Xi}(t') \rangle = 2 q \mathbf{1} \delta(t-t')$
with $q = \kbTe \mu_K = \kbTt \tilde{\mu}$ 
and we have defined the OLE mobility $\tilde{\mu} = \mu_K \alpha$ and the OLE temperature $\kbTt = \kbTe/\alpha$. This analysis shows that, in the overdamped limit,  it is not possible to distinguish between a renormalization of the mobility and a renormalization of the force, as long as the renormalized temperature is adjusted accordingly. Physically, this means that the separate effect of force renormalization -- as opposed to just temperature renormalization \cite{Szamel2021_efftemp} -- only becomes relevant if inertial effects are important. \js{As such, our results are most experimentally relevant for systems in which the constituents are of the meso- or macroscopic scale, such as active granular matter systems~\cite{des_granular3,weber_granular2,Narayan_granular1,kudrolli_granular4,Zik_1992,dauchot_granular1,dauchot_granular2,seguin_granular,Fiege_2012_granular} or systems of synthetic active robots~\cite{Soudeh_Lowen,leoni_robots,Tapia-Ignacio_2021}. However, in principle force renormalization can be distinguished from temperature renormalization for any system in which you can independently measure the kinetic temperature.}

The coarse-grained GLE for a single probe particle in a renormalized potential is equivalent to a GLE for an equilibrium system, therefore we cannot expect to find interesting nonequilibrium effects in such a model. However, the renormalization of temperature and forces has interesting consequences for systems of several interacting particles. We found that the kinetic temperature of the probe particle not only depends on the parameters of the bath such as density and activity level, but also on the properties of the probe, such as the size \cite{Shea_thesis}. A similar behavior can also be expected for the force renormalization factor $\alpha$. If probes of different size interact with each other, we therefore expect the forces to be rescaled differently, implying that the effective forces are non-reciprocal. This is a common signature of nonequilibrium \cite{Loos2020,Rosalba_NonRecip,Godrèche_2019,Kano_NonRecip,Tailleur_NonRecip,Ramaswamy_NonRecip,Golestanian_NonRecip}  whose consequences for our particular type of system yet remain to be explored.

Furthermore, we have seen that the GLE mapping with constant renormalization parameter $\alpha$ breaks down for very large external forces. If $\alpha$ depends on the force, the renormalization may turn a conservative force into a non-conservative drift force, i.e., it may no longer be possible to rewrite the renormalized force as a gradient of a renormalized potential. Developing coarse-grained methods for external forces beyond the linear response regime will be an important area for future research. Moreover, theoretical arguments (appendix \ref{sec:app_mem}) suggest that it might sometimes be necessary to modify the 2FDT by adding another force-dependent term, e.g.,  as in Eq.\ (\ref{eq:2fdt_mod}). In the case studies considered in the present work, the additional term was found to vanish; in general, it might be nonzero and generate true nonequilibrium signatures even in a single-particle model. 

It has been shown that a harmonically confined active particle exhibits different behavior from a harmonically confined passive particle~\cite{TrapABP_Brady,Franosch_Harmonic,Malakar_Harmonic}. Namely, the radial probability distribution of the active particle exhibits two `phases' which depend on the particle activity and the trap strength: a `passive' phase described by a Boltzmann-like distribution around the trap center, and an `active' phase described by a non-Boltzmann distribution which is peaked away from the trap center~\cite{Malakar_Harmonic}. Given the shared behaviors of a probe immersed in an active bath with an active particle itself~ \cite{steff,Callegari,Volpe}, it would be interesting to determine whether there exists an `active' phase probability distribution for a probe immersed in an active bath for other values of activity and spring potentials, and whether this can be described
by a a GLE with a modified 2FDT and/or a force-dependent $\alpha$-parameter.

\section*{Acknowledgements}
This work was funded by the Deutsche Forschungsgemeinschaft (DFG) via Grant 233630050, TRR 146, Project A3. Computations were carried out on the Mogon Computing Cluster at ZDV Mainz. We thank Martin Hanke, Niklas Bockius, and Thomas Speck for useful discussions.

\section*{Conflicts of interest}
There are no conflicts to declare.

\section*{Appendix}
\appendix

\section{Generalized equipartition theorem and modified 2FDT}
\label{sec:app_mem}

\fsnew{We consider an ensemble of trajectories of particles subject to an external, possibly position  dependent force $\mathbf{F}_\mathrm{ext}$. The underlying fine-grained system is taken to be in (possibly nonequilibrium) steady state (NESS). Our goal is to map this trajectory ensemble onto a GLE  that reproduces certain ensemble averages (denoted by $\langle ...\rangle$) of interest. Specifically, our target coarse-grained model is a stationary GLE\finaledits{~\cite{Shin_2010}} of the form}
     \begin{eqnarray}
        \label{eq:gle_V}
         M\dot{\mathbf{V}}(t) &=& \mathbf{f}(\fsnew{\mathbf{R}(t)})  - \int_{\jsnew{-\infty}}^{t} \text{d} s \: \mathbf{K}(\jsnew{t-s}) \mathbf{V}(s) + \bm{\Gamma}(t)
    \\    \dot{\mathbf{R}}(t) &=& \mathbf{V}(t). 
       \label{eq:gle_R}
     \end{eqnarray}
with $\langle \bm{\Gamma} (t)\rangle \equiv 0$, and $\langle \bm{\Gamma} (t) \bm{\Gamma}(t_0) \rangle \to 0$, 
$\bm{K}(\jsnew{t-t_0}) \to 0$ for $|t - t_0| \to \infty$. 
To keep things general, we describe the memory kernel $\mathbf{K}$ as a tensorial quantity $(K_{jk})$ here. We proceed as follows:
\begin{enumerate}

\item[(I)] We define tensorial correlation functions 
    \begin{align}
      \mathbf{C}_{\mathbf{V}}(\jsnew{t-t_0}) &= \langle \mathbf{V}(t)\mathbf{V}(t_0) \rangle, \\ \nonumber
      \mathbf{C}_{\mathbf{fV}}(\jsnew{t-t_0}) &= \langle \mathbf{f}(\fsnew{\mathbf{R}(t))} \:\mathbf{V}(t_0) \rangle
     \end{align}
and   tensorial quantities $\mathbf{\theta}_{\mathbf{w}}$
\begin{equation}
\label{eq:theta}
    \mathbf{\theta}_{\mathbf{w}} (\jsnew{t-t_0})=
    \langle \bm{\Gamma}(t) \: \mathbf{w}(t_0) \rangle 
      - \int_{\jsnew{-\infty}}^{t_0} \mathrm{d}s \: \mathbf{K}(\jsnew{t-s}) \: \langle \mathbf{V}(s) \: \mathbf{w}(t_0) \rangle.
\end{equation}
where $\mathbf{w}$ is a placeholder for $\mathbf{w}= \mathbf{V}$, $\mathbf{f}$, or $\mathbf{R}$.
\fsnew{We consider NESS situations, hence the correlation functions $\mathbf{C}_{\mathbf{V}}$,  $\mathbf{C}_{\mathbf{fV}}$ only depend on the time difference $t-t_0$. Furthermore, since we map onto a stationary GLE,  this also holds for correlation functions $\langle \bm{\Gamma}(t) \: \mathbf{w}(t_0) \rangle$ involving noise. Note that the latter would no longer be true if we chose to map onto a non-stationary GLE model (Eq. (\ref{eq:gle})) with finite start time $T$. However, the arguments and derivations below remain valid if we replace lower  bounds $(- \infty)$ in time integrals by $T$ and $\mathbf{\theta}_{\mathbf{w}} ( t-t_0)$ by    
\begin{displaymath}
    \mathbf{\theta}_{\mathbf{w}} (t,t_0)=
    \langle \bm{\Gamma}(t) \: \mathbf{w}(t_0) \rangle 
      - \int_T^{t_0} \mathrm{d}s \: \mathbf{K}(t-s) \: \langle \mathbf{V}(s) \: \mathbf{w}(t_0) \rangle.
\end{displaymath}
}

By multiplying the GLE, (\ref{eq:gle_V}), with $\mathbf{V}(t_0)$ and taking the \fsnew{trajectory ensemble} average, we obtain the equation
\begin{eqnarray}
    \label{eq:volterra_with_rest}
    M \langle \dot{\mathbf{V}}(t) \: \mathbf{V}(t_0) \rangle  
    &=& \langle \mathbf{f}(\fsnew{\mathbf{R}(t)}) \: \mathbf{V}(t_0) \rangle
       \\ && \nonumber - \int_{t_0}^t \mathrm{d}s \: \mathbf{K}(\jsnew{t-s}) \: \langle \mathbf{V}(s) \: \mathbf{V}(t_0) \rangle
       + \mathbf{\theta}_{\mathbf{V}}(\jsnew{t-t_0}).
\end{eqnarray}

In order to define the memory kernel unambiguously, we impose the Volterra equation,
  \begin{equation}
    \label{eq:volterra}
    M \frac{\mathrm{d}}{\mathrm{d}t} \mathbf{C}_{\mathbf{V}}(\jsnew{t-t_0})
    = \mathbf{C}_{\mathbf{fV}}(\jsnew{t-t_0}) 
       - \int_{t_0}^t \mathrm{d}s \: \mathbf{K}(\jsnew{t-s}) \: \mathbf{C}_{\mathbf{V}}(\jsnew{s-t_0})
\end{equation}
(assuming that it can be solved with respect to ${\mathbf{K}}$\js{)}, 
which is equivalent to imposing $\theta_{\mathbf{V},jk}(\jsnew{t-t_0}) \equiv 0$ (noise orthogonality with respect to velocity \cite{Shin_2010,wip}).
It is then convenient to define the one-sided Fourier transform
 \begin{eqnarray}
 \mathcal{F} [\chi] = \widetilde{\chi}(\omega) 
  = \int_{0}^{\infty} \chi(t) \: \textrm{e}^{\textrm{i} \omega (t)} \textrm{d}t
 \end{eqnarray}
  for general functions $\chi(t )$. Specifically, $\widetilde{\mathbf{K}}(\omega = 0) = \int_0^\infty \textrm{d}\tau \: \mathbf{K}(\tau)=\bm{\mu}_K^{-1}$  is the inverse tensorial GLE mobility as defined in Eq. (\ref{eq:diffcoeff_k}).
  \jsnew{We consider the GLE mapping to be successful if $\bm{\mu}_K^{-1}$ exists and does not diverge. Additionally, for a successful mapping,} the limit $\lim_{t \to \infty} \mathbf{K}(t)$ \fsnew{must exist}, which implies $\lim_{t \to \infty} \mathbf{K}(t) = 0$.
  We also note that $\widetilde{\mathbf{C}}_V(\omega = 0) = \int_0^\infty \textrm{d}\tau \: \mathbf{C}_V(\tau) =: \mathbf{D}$
  is the diffusion tensor for diffusing particles. 

Applying the Fourier transform to the Volterra equation, we obtain 
 \begin{align}
 \label{eq:volterra_ft}
 -\textrm{i} \omega M \widetilde{\mathbf{C}}_{\mathbf{V}}(\omega) = \widetilde{\mathbf{C}}_{\mathbf{fV}}(\omega)
        - \widetilde{\mathbf{K}}(\omega) \widetilde{\mathbf{C}}_{\mathbf{V}}(\omega) 
        + M \langle \mathbf{V} \mathbf{V} \rangle
 \end{align}
 where we have used standard relations for the Fourier transform and $\mathbf{C}_\mathbf{V}(0) = \langle \mathbf{V} \mathbf{V} \rangle$.

 \item[(II)]
We first discuss the case of a free particle with $U^{\textrm{eff}} \equiv 0$.
 In that case, the Volterra equation in the limit $\omega \to 0$ reads $ M \langle \mathbf{V} \mathbf{V} \rangle = \widetilde{\mathbf{K}}(\omega=0) \: \widetilde{\mathbf{C}}_V(\omega=0)$ and hence we obtain
 a generalized Einstein relation between the mobility and diffusion tensors,
 \begin{equation}
 \label{eq:einstein}
    \mathbf{D}= \bm{\mu}_{\textrm{K}} \: M \langle \mathbf{V} \mathbf{V} \rangle
 \end{equation}
 In isotropic systems, one has $ M \langle \mathbf{V} \mathbf{V} \rangle = \bm{1} \: \kbTe $, 
 $\mathbf{D} = \bm{1} \:D  $, and $\bm{\mu}_{\textrm{K}} = \bm{1} \: \mu_{\textrm{K}}  $, and the
 generalized Einstein relation simplifies to the more common form $D = \kbTe \: \mu_{\textrm{K}}$. \jsnew{From Eq.~\eqref{eq:volterra_ft}, we can deduce that $\tilde{K}(0)$ is not bounded if $\tilde{C}_{\mathbf{V}}(0)=0$. In this case, both the diffusion constant and the mobility are zero. }

 \item[(III)] 
  Next we consider a confined particle subject to a confining potential $U(\mathbf{R})$, such that $\langle \mathbf{R}^2 \rangle < \infty$, e.g., a particle in an external potential with infinite barrier height. It does not diffuse, hence $\widetilde{\mathbf{C}}_V(\omega = 0)= 0$. Since $\widehat{\mathbf{K}}(\omega =0)$ is finite
(one of our basic assumptions) and $\widetilde{\mathbf{C}}_V(\omega = 0)= 0$, the memory
term in Eq.\ (\ref{eq:volterra_ft}) vanishes, and so does the term $- i \omega M \widetilde{\mathbf{C}}_V$ 
on the left hand side of the equation. Thus the Volterra equation only has a solution if
 $M \langle \mathbf{V} \mathbf{V} \rangle =- \lim_{\omega \to 0} \widetilde{\mathbf{C}}_{\mathbf{fV}}(\omega)$, which implies
 \begin{align}
 \nonumber
  M \langle \mathbf{V} \mathbf{V} \rangle &=  \lim_{\omega \to 0}\mathcal{F} [\langle \mathbf{V}(0) \: 
     \nabla U^\text{eff}(\mathbf{R}(t)) \rangle  ] \\  \nonumber &
   =  \int_0^\infty \langle \mathbf{V}(0) \: \nabla U^\text{eff}(\mathbf{R}(t)) \rangle \: \mathrm{d}t \\ \nonumber
   &=  \int_{0}^{\infty} \langle  \mathbf{V}(-t) \:\nabla U^\text{eff}(\mathbf{R}(0)) \rangle  
  \: \mathrm{d}t \\  \nonumber & 
  = -\int_{0}^{\infty} \frac{\mathrm{d}}{\mathrm{d}t} 
     \langle \mathbf{R}(-t) \: \nabla U^\text{eff}(\mathbf{R}(0))   \rangle  \: \mathrm{d}t  \\   &
  = \langle   \mathbf{R}(0)\: \nabla U^\text{eff}(\mathbf{R}(0)) \rangle 
  =    \langle   \mathbf{R}\: \nabla U^\text{eff} \rangle\jsnew{,}         
 \end{align}
\\
 \jsnew{which is listed as Eq.~\eqref{eq:equipartition} in the main text. We note that }\fsnew{this condition is equivalent to requesting $\theta_\mathbf{R}(0) = 0$. This can be seen by multiplying Eq.~\eqref{eq:gle_V} with $\mathbf{R}(t)$, inserting  $\mathbf{f} = - \nabla U^\text{eff}$, taking the trajectory ensemble average, and using  $0=\frac{\textrm{d}}{\textrm{d} t}\langle \mathbf{R}\mathbf{V}\rangle = \langle \mathbf{R} \dot{\mathbf{V}} \rangle+\langle \mathbf{V} \mathbf{V}\rangle$. 
 }
 
 
 Applying the general result  \eqref{eq:equipartition}  to the harmonic oscillator potential,
 $ U^\text{eff}(\mathbf{R}) = \frac{1}{2} \tilde{k} \: \mathbf{R}^2$, we find that the
 effective spring constant must fulfill 
 $\tilde{k} = M \langle \mathbf{V}^2 \rangle/{\langle \mathbf{R}^2 \rangle} = \kbTe/\langle x^2 \rangle$,
 which corresponds to the renormalization of the spring constant in the main text.

\item[(IV)] Finally, we analyze the implications of (\ref{eq:volterra}) for the fluctuations of the stochastic noise $\bm{\Gamma}(t)$. We have analyzed the force free case in Ref. \cite{wip} and shown that the Volterra equation then enforces a generalized 2FDT 
\begin{equation}
    \label{eq:2fdt}
    \langle \bm{\Gamma}(t) \bm{\Gamma}(t_0) \rangle  = 
    M \mathbf{K}(\jsnew{t-t_0}) \: \mathbf{C}_{\mathbf{V}}(0), 
\end{equation}
(Eq.\ (\ref{eq:2fdt_gen}) in the main text). Here we use Eqs. (\ref{eq:gle_V}) and (\ref{eq:volterra_with_rest}) as a starting point and follow the same steps \jsnew{as} Ref. \cite{wip} (Appendix A) to derive a corresponding equation in the presence of a drift force.
First, 
\fsnew{We write Eq.~\eqref{eq:gle_V} in terms of a time variable $t_0$}
\jsnew{and then multiply this equation by $\bm{\Gamma}(t)$. We then take the ensemble average to obtain that}
\begin{eqnarray}
\label{eq:2fdt_der1}
    \langle \Gamma_j(t) \Gamma_k(t_0) \rangle
    &=& M \langle \Gamma_j(t) \dot{V}_k(t_0) \rangle
    - \langle \Gamma_j(t) f_k(t_0)\rangle \\
    && \nonumber + \int_{\jsnew{-\infty}}^{t_0} \mathrm{d}s \: K_{kl}(\jsnew{t_0-s}) \langle \Gamma_j(t) V_l(s) \rangle.
\end{eqnarray}
Here, $\mathbf{f}(t)$ stands for $\mathbf{f}(t)=\mathbf{f}(\fsnew{\mathbf{R}(t)})$, and we
use the Einstein summation convention. \js{For the remainder of the derivation, we will indicate the insertion of previous results with brackets $\mathbf{\Big[...\mathbf{\Big]}}$.}
With the definition of  $\mathbf{\theta}_{\mathbf{V}}$ (Eq.\ (\ref{eq:theta})), we
can express the first term on the right hand side (r.h.s.) of (\ref{eq:2fdt_der1}) as
\begin{eqnarray}
\label{eq:2fdt_der2}
    \lefteqn{ M \langle \Gamma_j(t) \dot{V}_k(t_0) \rangle
     = M \frac{\mathrm{d}}{\mathrm{d} t_0} \langle \Gamma_j(t) {V}_k(t_0) \rangle} \\ \nonumber
     &=&  
      M \frac{\mathrm{d}}{\mathrm{d} t_0}  \js{\mathbf{\Big[}}
     \theta_{\mathbf{V},jk} (\jsnew{t-t_0}) + 
     \int_{\jsnew{-\infty}}^{t_0} \mathrm{d}s \: K_{jl}(\jsnew{t-s}) 
      \fsnew{\mathbf{C}_{\mathbf{V}, lk}( s-t_0)}
      \js{\mathbf{\Big]}} 
     \\ \nonumber &=&
     M \frac{\mathrm{d}}{\mathrm{d} t_0} \theta_{\mathbf{V},jk} (\jsnew{t-t_0})
     + M K_{jl}(\jsnew{t-t_0}) 
      \fsnew{\mathbf{C}_{\mathbf{V}, lk}(0)}
     \\ \nonumber
     && + \: M \int_{\jsnew{-\infty}}^{t_0} \mathrm{d}s \: K_{jl}(\jsnew{t-s}) 
      \fsnew{\frac{\mathrm{d}}{\mathrm{d} t_0}\mathbf{C}_{\mathbf{V}, lk}(s-t_0)}
     \js{.}
\end{eqnarray}
\js{ We can additionally re-express the last term on the r.h.s. of Eq.~(\ref{eq:2fdt_der1}) using the definition of  $\mathbf{\theta}_{\mathbf{V}}$ given in Eq.\ (\ref{eq:theta}):} 
\begin{eqnarray}
\label{eq:2fdt_der3} 
     \lefteqn{\int_{\jsnew{-\infty}}^{t_0} \!\!\! \mathrm{d}s \: K_{kl}(\jsnew{t_0-s}) \langle \Gamma_j(t) V_l(s) \rangle} 
     \\ &=& \nonumber \!\!\!\!\!
     \int_{\jsnew{-\infty}}^{t_0} \!\!\! \mathrm{d}s \: K_{kl}(\jsnew{t_0-s}) \times
\\ \nonumber && 
     \js{\mathbf{\Big[}} \theta_{\mathbf{V},jl} (\jsnew{t-s})   
     + \int_{\jsnew{-\infty}}^{s} \mathrm{d}s' \: K_{ji}(\jsnew{t-s'}) 
     \fsnew{\mathbf{C}_{\mathbf{V}, il}( s'-s)}
     \js{\mathbf{\Big]}} 
\end{eqnarray}
Now, \fsnew{using
the symmetry relation  $\mathbf{C}_{\mathbf{V},ij}(t-t') = \mathbf{C}_{\mathbf{V},ji}(t' -t )$},
we rewrite the \fsnew{second} term on the r.h.s.
of (\ref{eq:2fdt_der3}) as
\begin{eqnarray}
\label{eq:2fdt_der4}
     \lefteqn{\int_{\jsnew{-\infty}}^{t_0} \!\!\!  \mathrm{d}s \: 
     \int_{-\infty}^{s} \!\!\! \mathrm{d}s' 
     \: K_{kl}(\jsnew{t_0-s}) \:  K_{ji}(\jsnew{t-s'}) \: 
          \fsnew{\mathbf{C}_{\mathbf{V}, il}( s'-s)}     
     } 
     \\ &=& \nonumber 
     \int_{\jsnew{-\infty}}^{t_0} \!\!\!  \mathrm{d}s' \:  K_{ji}(\jsnew{t-s'}) \int_{s'}^{t_0} \!\!\! \mathrm{d}s
     \: K_{kl}(\jsnew{t_0-s}) \: 
     \fsnew{\mathbf{C}_{\mathbf{V}, li}( s-s')}
     \qquad \qquad
     \\ &=& \nonumber 
     \int_{\jsnew{-\infty}}^{t_0} \!\!\! \mathrm{d}s' \:  K_{ji}(\jsnew{t-s'})
     \js{\mathbf{\Big[}} - M 
        \fsnew{\frac{\mathrm{d}}{\mathrm{d} t_0}\mathbf{C}_{\mathbf{V}, ik}(s'-t_0)}
     \\  \nonumber && \quad
         + \: \fsnew{\mathbf{C}_{\mathbf{fV}, ki}( t_0-s')}
     + \theta_{\mathbf{V},ki}(\jsnew{t_0-s'}) \js{\mathbf{\Big]}.}
\end{eqnarray}
\fsnew{In the last step, we have inserted Eq.\ (\ref{eq:volterra_with_rest}) and used
the symmetry relation  for $\mathbf{C}_{\mathbf{V} }$} once more.
Combining (\ref{eq:2fdt_der1})-(\ref{eq:2fdt_der4}), we obtain the following general tensorial expression for the noise fluctuations:
\begin{eqnarray}
    \lefteqn{\langle \bm{\Gamma}(t) \bm{\Gamma}(t_0) \rangle  = 
    M \:\mathbf{K}(\jsnew{t-t_0}) \: \mathbf{C}_{\mathbf{V}}(\jsnew{0}) - \mathbf{\theta}_{\mathbf{f}}(\jsnew{t-t_0})} 
    \\ && \nonumber     
     + \: M \frac{\mathrm{d}}{\mathrm{d} t_0} \mathbf{\theta}_{\mathbf{V}}(\jsnew{t-t_0})
      \\  && \nonumber 
     + \int_{\jsnew{-\infty}}^{t_0} \!\!\! \mathrm{d}s \: \Big( \mathbf{K}(\jsnew{t-s}) \mathbf{\theta}_{\mathbf{V}}^T(\jsnew{t_0- s}) 
       + \: \mathbf{\theta}_{\mathbf{V}}(\jsnew{t-s}) \mathbf{K}^T(\jsnew{t_0-s})  \Big).
\end{eqnarray}
The last \fsnew{three} terms vanish \js{because} $\mathbf{\theta}_{\mathbf{V}} \equiv 0$. However, the generalized 2FDT, Eq.\ (\ref{eq:2fdt}), is only recovered if one has, in addition, $\mathbf{\theta}_{\mathbf{f}}(\jsnew{t-t_0}) \equiv 0$ (noise orthogonality with respect to the force). As discussed in  Ref.\ \cite{wip}, this condition can be enforced if the force is constant or only depends explicitly on time, as in the example of the constant drag force. On the other hand, it may be broken if the force depends on the position of the particle \cite{Glatzel_2021, Schilling_2022}. In such cases,
the 2FDT is modified according to
\begin{equation}
    \label{eq:2fdt_modified}
    \langle \bm{\Gamma}(t) \bm{\Gamma}(t_0) \rangle  = 
    M \mathbf{K}(\jsnew{t-t_0}) \: \mathbf{C}_{\mathbf{V}}(\jsnew{0}) - \mathbf{\theta}_{\mathbf{f}}(\jsnew{t-t_0}).
\end{equation}

We now specifically consider the case of the (possibly renormalized) harmonic potential, where the drift force can be expressed as $\mathbf{f} = - \tilde{k} \mathbf{R}$. Then we have $\mathbf{\theta}_{\mathbf{f}}(\jsnew{t-t_0})^{\textrm{harm.}} = - \tilde{k} \mathbf{\theta}_{\mathbf{R}}(\jsnew{t-t_0})$. \fsnew{$\mathbf{\theta}_{\mathbf{R}}( t-t_0)$ can be calculated explicitly as follows: We first take the derivative of $\mathbf{\theta}_{\mathbf{R}}(\jsnew{t-t_0})$ with respect to $t_0$, and obtain
\begin{align}
\nonumber
    \frac{\textrm{d}}{\textrm{d}t_0} \mathbf{\theta}_{\mathbf{R}}(\jsnew{t-t_0}) 
   &= \mathbf{\theta}_{\mathbf{V}}(\jsnew{t-t_0})
    -  \mathbf{K}(\jsnew{t-t_0}) \: \langle \mathbf{V}(t_0) \: \mathbf{R}(t_0) \rangle
   \\ &=  -\: \mathbf{K}(\jsnew{t-t_0}) \: \fsnew{\langle \mathbf{V} \: \mathbf{R} \rangle}.
\end{align}
where we have used $\mathbf{\theta}_{\mathbf{V}}(\jsnew{t-t_0})\equiv 0$.  Integrating over $t_0$ and using $\mathbf{\theta}_{\mathbf{R}} (\jsnew{t-t_0})\to0$ at $(t-t_0) \to \infty$ gives an  expression for $\mathbf{\theta}_{\mathbf{R}}$, which is generally valid:
\begin{equation}
\label{eq:NER}
\mathbf{\theta}_{\mathbf{R}}(\tau) = 
 - \mathbf{I}(\tau) \: \langle \mathbf{V}  \mathbf{R} \rangle 
 \quad \mbox{with} \quad
       \mathbf{I}(\tau)  =\int_{\tau}^{\infty} \textrm{d}s \: \mathbf{K}(s)
\end{equation}
} 

\fsnew{Inserting this result into Eq. (\ref{eq:2fdt_modified}),} 
 we finally obtain the following expression for the modified 2FDT in the presence of a harmonic potential: 
\begin{equation}
\label{eq:modified_2fdt} 
  \langle \bm{\Gamma}(t) \bm{\Gamma}(t_0) \rangle
     = M \mathbf{K}(t-t_0) \: \langle \mathbf{V} \mathbf{V} \rangle 
       + \tilde{k} \:  \mathbf{I}(t-t_0) \: \langle \mathbf{V}  \mathbf{R} \rangle  
\end{equation}

At equilibrium, we have $\langle \mathbf{V} \mathbf{R} \rangle = 0$ due to time reversal symmetry. Therefore, the last term vanishes and one recovers the force-free generalized 2FDT, Eq.\ (\ref{eq:2fdt_gen}) in the main text. In active systems, this is not necessarily the case.
For example, active particles have a tendency to accumulate at boundaries in a state where the active velocity points towards the boundary, which creates a clear correlation between position and velocity.
If such a correlation were present for a harmonically trapped particle,
omitting the last term in Eq. (\ref{eq:modified_2fdt}) in the corresponding coarse-grained GLE  would effectively map the system onto an equilibrium system, where $\langle \mathbf{V} \mathbf{R} \rangle = 0$ by construction. In such cases, we can only recover $\langle \mathbf{V} \mathbf{R} \rangle \neq 0$ by using the full modified 2FDT. \js{We note that it may be possible to define projection operators which formally lead to other definitions of the memory kernel and thus other FDTs in which the $\langle \mathbf{V} \mathbf{R} \rangle$ term may or may not be present. Here, however, we do not apply
projection operator techniques, but rather analyze the GLE
under the sole assumption that the memory is given by the Volterra equation. In this case, provided that the system is not at equilibrium, the $\langle \mathbf{V} \mathbf{R} \rangle$ term cannot generally be eliminated.}

\end{enumerate}

Numerically, however, we found the additional term in the 2FDT to be negligible in the systems considered in the present work. The value of $\langle \mathbf{V} \mathbf{R} \rangle$ was always zero within the error. 
\fsnew{We note that, due to Eq.~(\ref{eq:NER}), this also implies $\mathbf{\theta}_{\mathbf{R}}(\jsnew{t-t_0})\equiv 0$ 
(noise orthogonality with respect to the position).}  On the other hand, the force renormalization in an active bath was substantial, as reported in the main paper, and moreover, the correct implementation of the force renormalization was crucial for the memory reconstruction, see appendix \ref{sec:app_memory_reconstruction}.

%

\section{Renormalization constant as a function of $\hat{F}_0$}
\label{app:alpha_f0}
\begin{figure}
  \centering
  \includegraphics[width=.7\linewidth]{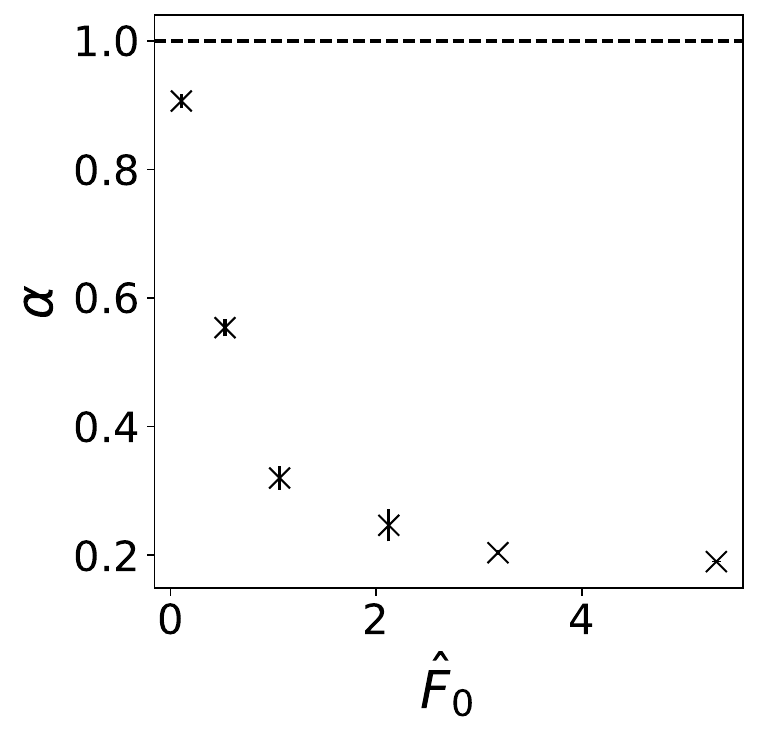}  
\caption{\js{The renormalization constant $\alpha$ as a function of the bath activity $\hat{F}_0$. The dashed black line shows $\alpha=1$, what we would expect for a passive bath. The bath has average density $\rho=0.4\sigma^{-3}$ and the probe has a radius $R_p=3.0\sigma$.}}
\label{fig:alpha_f0}
\end{figure}

\js{In the main part of this manuscript, we focus on the system of a probe either immersed in a passive bath or immersed in an active bath with activity $\hat{F}_0=5.3$. This same type of analysis can be performed for baths of different activities, i.e. with different activities $\hat{F}_0$. In particular, the necessity of force renormalization is also applicable to baths with different activities. To illustrate this point, we calculated the value the renormalization constant $\alpha$ for baths with different activities $\hat{F}_0$. We calculated $\alpha$ using Eq.~\ref{eq:eff_drift_force} and the fact that, effectively, $\alpha=\hat{f}_x/\hat{F}_{\textrm{ext}}$. For each value of $\hat{F}_0$, we calculated $\alpha$ using two values of $F_\mathrm{ext}$ where $0.1<\mathrm{Pe}<0.5$. The results are shown in Fig.~\ref{fig:alpha_f0}. From Fig.~\ref{fig:alpha_f0}, we can see that, even for low activity baths, a renormalized force is necessary.} 

\section{Numerical reconstruction of the memory kernel of a harmonically trapped particle}
\label{sec:app_memory_reconstruction}

Our starting point is a scalar and steady-state version of the Volterra equation, (\ref{eq:volterra}): 
  \begin{equation}
    \label{eq:volterra_stationary}
    M \dot{C}_V(t-t_0) =  \langle \mathbf{f}(t) \cdot \mathbf{V}(t_0) \rangle - \int_{t_0}^t \mathrm{d}s \: K(t-s)\: C_V(s),
\end{equation}
with $C_V(t-t_0) = \langle \mathbf{V}(t) \cdot \mathbf{V}(t_0) \rangle$ (see main text). We insert the renormalized
harmonic spring force, $\mathbf{f} = - \tilde{k} \: \mathbf{R}$, take the derivative with respect to $t$, 
and use $M \dot{\mathbf{V}} = \mathbf{F}$.
Furthermore, we exploit the relations 
\begin{align}
\nonumber
 \frac{\textrm{d}}{\textrm{d} t} \langle {\mathbf{W}}(t) \cdot \mathbf{V}(t_0) \rangle 
&= \frac{\textrm{d}}{\textrm{d} t} \langle {\mathbf{W}}(0) \cdot \mathbf{V}(t_0-t) \rangle \\ \nonumber
&= -  \langle \mathbf{W}(0) \cdot \dot{\mathbf{V}}(t_0-t) \rangle \\ \nonumber
&= -  \langle \mathbf{W}(t) \cdot \dot{\mathbf{V}}(t_0) \rangle 
\end{align}
for   $\mathbf{W} = \mathbf{F}$ and $\mathbf{W} = \mathbf{R}$ and the relation 
\begin{align}
\nonumber
\frac{\textrm{d}}{\textrm{d} t} \int_{t_0}^t \mathrm{d}s \: K(t-s)\: C_V(s) 
&=\frac{\textrm{d}}{\textrm{d} t} \int_{t_0}^t \mathrm{d}s \: C_V(t-s)\: K(s) 
\\ \nonumber &= C_V(t_0) K(t) + \int_{t_0}^t \mathrm{d}s \: K(s)\: \dot{C}_V(t-s)
\end{align} 
Setting $t_0=0$, this gives the Volterra equation of the second kind quoted in the main text, 
\begin{equation}
\label{eq:volterra2}
C_F(t) = M \tilde{k} \:C_V(t) + \int_0^t \mathrm{d}s \: K(s) C_{FV}(t-s) + M K(t) C_V(0)
\end{equation}
with $C_F(t) = \langle \mathbf{F}(t) \cdot \mathbf{F}(0) \rangle$ an $C_{FV}(t) = \langle \mathbf{F}(t) \cdot \mathbf{V}(0) \rangle$. Eq. (\ref{eq:volterra2}) is inverted using ~\cite{Shin_2010}: 
\begin{align}
\label{eq:step_mem_harm}
K(m \Delta t)&=\left[C_V(0)+\frac{\Delta t}{2 M}C_{FV}(0)\right]^{-1}\Bigg[\frac{1}{M} C_F(m\Delta t)-\tilde{k} \: C_V(m\Delta t) \nonumber \\ \nonumber & -\frac{\Delta t}{M} \sum^{m-1}_{n=1}  C_{FV}((m-n)\Delta t)K(n\Delta t) \nonumber  - \frac{\Delta t}{2M} C_{FV}(m \Delta t) K(0)\Bigg] \\  ,
\end{align}
with the initial condition $K(0)=\frac{1}{M}\frac{C_F(0)}{C_V(0)}-\tilde{k}$. 

\begin{figure}
  \centering
  \includegraphics[width=1.\linewidth]{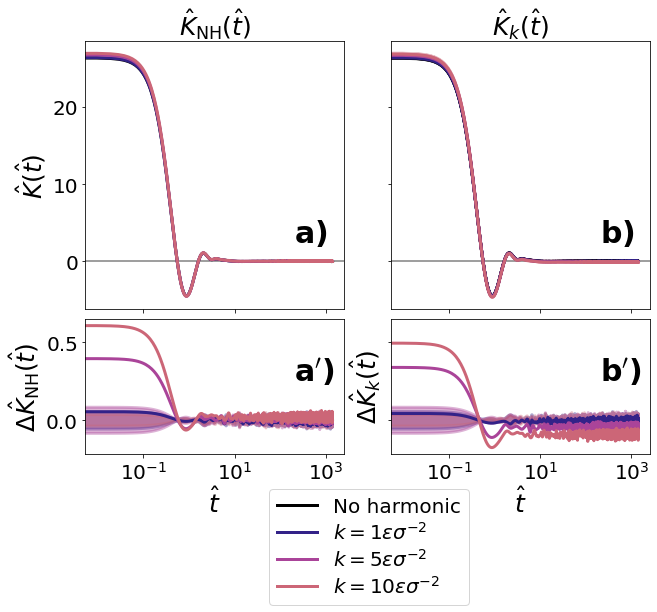}  
\caption{Memory kernel of a probe immersed in an active bath ($\hat{F}_0=5.3$) and trapped by a harmonic potential with strength $k$, where the memory kernel is calculated \js{a) without including the harmonic potential, b)} using the imposed value of the harmonic constraint $k$ rather than the renormalized value $\Tilde{k}$. Light shading indicates error bars. \js{Smaller graphs (a$^\prime$) and b$^\prime$)) show} the difference between the memory kernel of a free \js{probe} and a trapped probe\js{, for the memory kernels shown in the graph immediately above}. The bath has average density $\rho=0.4\sigma^{-3}$ and the probe has a radius $R_p=3.0\sigma$. }
\label{fig:app_harm_mem}
\end{figure}
 In Section~\ref{sec:harm_mk}, we used the the renormalized harmonic constraint $\Tilde{k}$ in Eq.~\ref{eq:step_mem_harm} to solve for the probe memory kernel based on our findings in Section~\ref{sec:harm_pos}. Additionally, we tried solving the memory kernel without a harmonic constraint ($\hat{K}_\mathrm{NH}(\hat{t})$) as well as with the imposed harmonic constraint $k$ ($\hat{K}_k(\hat{t})$). The results are shown in Figs.~\ref{fig:app_harm_mem}a) and b) respectively. Although Figs.~\ref{fig:app_harm_mem}a) and b) both look qualitatively the same as Fig.~\ref{fig:harm_vv}d), comparing the \js{Figs.~\ref{fig:app_harm_mem}a$^\prime$) and b$^\prime$)} with \js{Fig.~\ref{fig:harm_vv}d$^\prime$)} reveals that they are not the same. In \js{Fig.~\ref{fig:harm_vv}d$^\prime$)}, where we define $\hat{K}(\hat{t})$ with $\Tilde{k}$, the difference between the memory kernel of the constrained probe and that of a free probe goes to zero. However, this difference converges to a finite value at late times in the case that we either neglect the harmonic constraint (\js{Fig.~\ref{fig:app_harm_mem}a$^\prime$)}) or we define $\hat{K}(\hat{t})$ with $k$ (\js{Fig.~\ref{fig:app_harm_mem}b$^\prime$)}). This finding is similar to the renormalization/linearization within the Mori-Zwanzig formalism in the presence of non-linear external potentials \cite{Zwanzig_Non,jungjung2023}. We know that the memory kernel of the free probe goes to zero in the long time limit\cite{ME}. Therefore, the fact that $\hat{K}_\mathrm{NH}(\hat{t})$ and $\Delta\hat{K}_k(\hat{t})$ go to finite values, indicates that they do not approach zero in the long time limit as they should. Comparing \js{Fig.~\ref{fig:app_harm_mem}a$^\prime$)} to that of \js{Fig.~\ref{fig:app_harm_mem}b$^\prime$)}, we see that $\hat{K}_\mathrm{NH}(\hat{t})$ approaches a smaller (in terms of magnitude) finite value than $\Delta\hat{K}_k(\hat{t})$. This can be understood given that, in our case, the value of $\Tilde{k}$ is closer to zero than to $k$. We conclude that consistent force renormalization is essential for the successful reconstruction of a memory kernel via the Volterra equation formalism, i.e., the successful derivation of a GLE that satisfies the 2FDT.

\section{Velocity distribution of the harmonically trapped probe}
\label{sec:app_veldist}

\begin{figure}
  \centering
  \includegraphics[width=1.\linewidth]{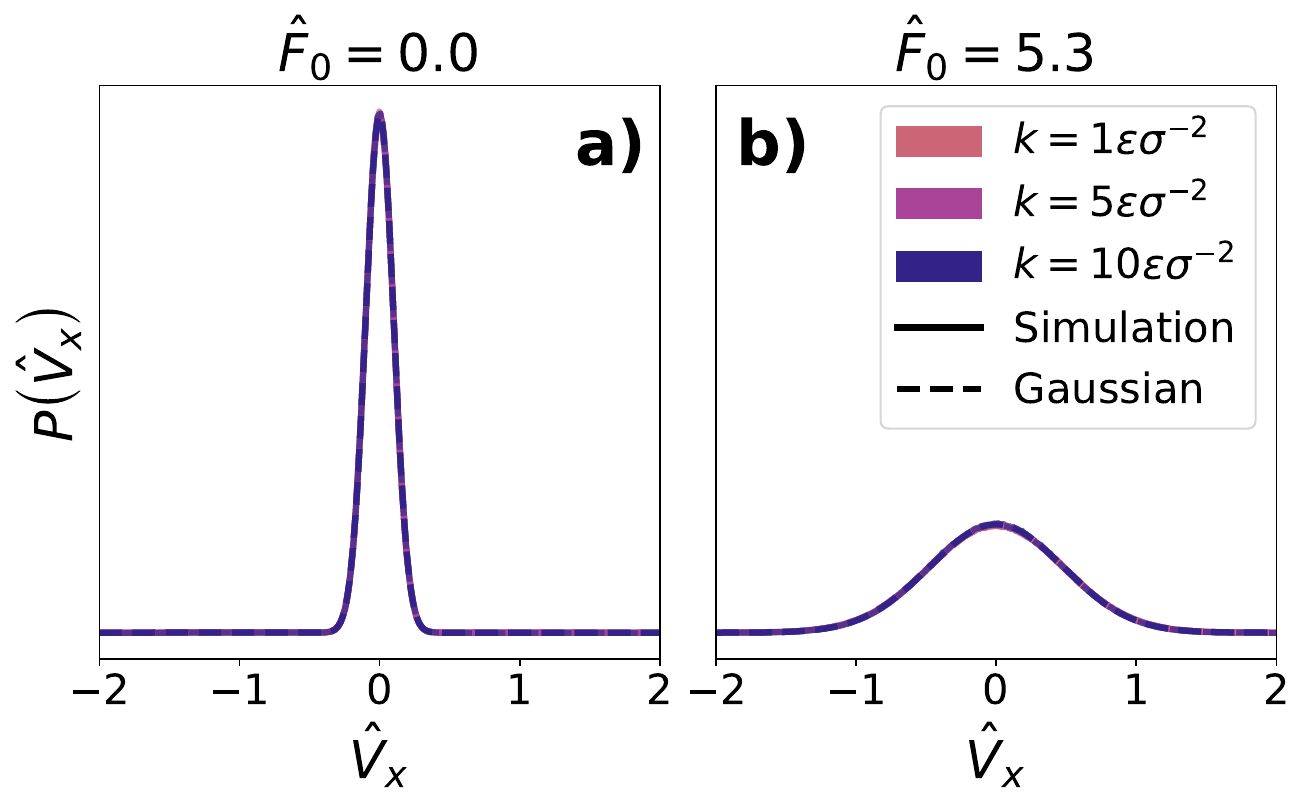}  
\caption{Velocity distribution of the harmonically trapped probe in a bath of density $\rho=0.4\sigma^{-3}$. \textbf{a)} The velocity distribution for a passive bath. \textbf{b)} The velocity distribution for an active bath  with activity $\hat{F}_0=5.3$. The solid lines show simulation data, whereas the dotted lines show zero-centered Gaussian distributions with the same standard deviation. In both \textbf{a)} and \textbf{b)} the Gaussian curves overlap the simulation data.}
\label{fig:app_veldist}
\end{figure}

We saw in Fig.~\ref{fig:harm_vv}b) that the kinetic temperature of the probe is dependent on the strength of the harmonic potential, $k$. However, within our range of $k$ values this effect is marginal. In fact, as can be see in Fig.~\ref{fig:app_veldist}, the velocity distributions of the probe do not show any apparent deviations across the different values of $k$ tested. Although there is a slight quantitative difference in the mean squared velocity (as seen from the different initial values of the VACF in Fig.~\ref{fig:harm_vv}b)), this difference is too small to be qualitatively noticeable.

\section{Peclet number as a function of drag force}
\label{sec:app_pefext}
\begin{figure}
  \centering
  \includegraphics[width=1.\linewidth]{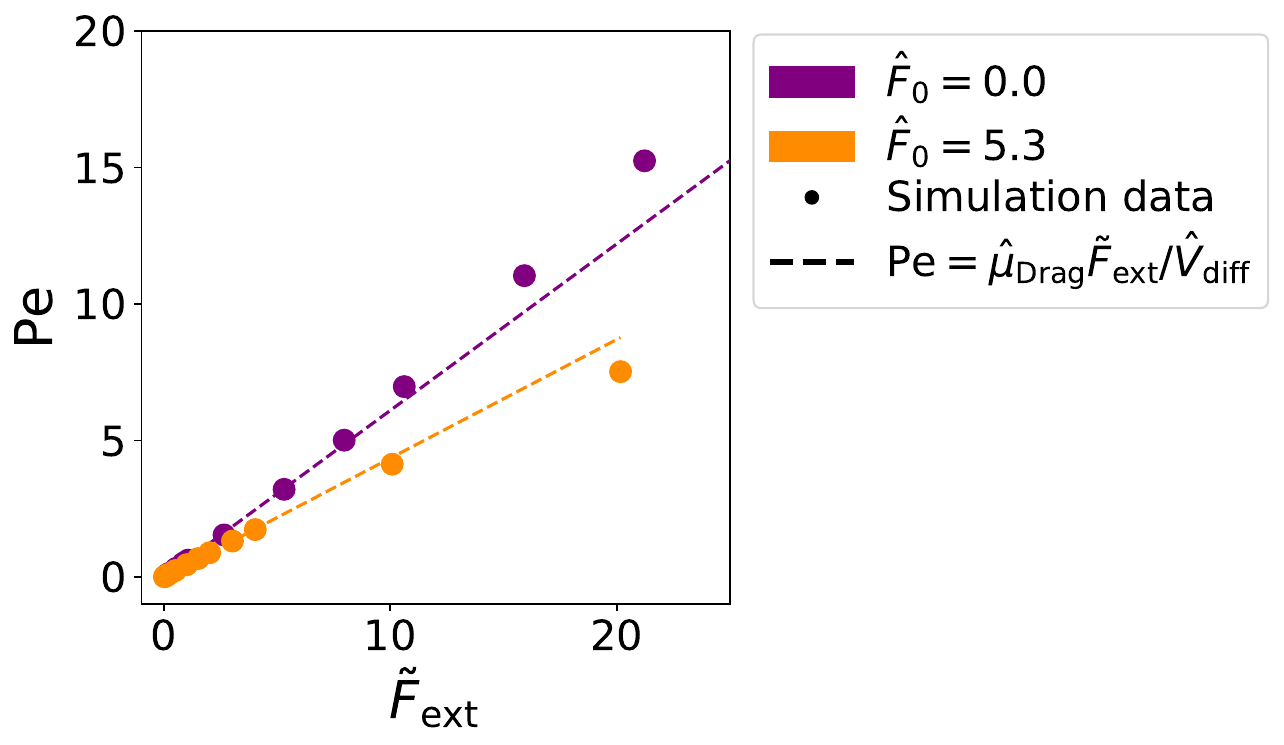}  
\caption{Peclet number ($\mathrm{Pe}$) as a function of the renormalized external drag force ($\Tilde{F}_\mathrm{ext}$). The dotted lines show $\mathrm{Pe}=\mu_\mathrm{Drag}\Tilde{F}_\mathrm{ext}/\hat{V}_\mathrm{diff}$ in the linear response regime, where $\mu_\mathrm{Drag}$ is constant.}
\label{fig:app_pefext}
\end{figure}

In Section~\ref{sec:mobility}, we define the Peclet number $\mathrm{Pe}=\langle \hat{V}_x \rangle/\hat{V}_\mathrm{diff}$ for the dragged probe, with $\hat{V}_\mathrm{diff}=D_\mathrm{eff}\sqrt{{m}/{\kbT}}/R_p$. Given that the average force exerted on the probe can be related to its average velocity through Eq.~\eqref{eq:diffcoeff_drag}, we can rewrite the Peclet number in terms of $\Tilde{F}_\mathrm{ext}$ as $\mathrm{Pe}=\mu_\mathrm{Drag}\Tilde{F}_\mathrm{ext}/\hat{V}_\mathrm{diff}$. We found in Section~\ref{sec:mobility} that, for both a probe immersed in an passive and an active bath, $\mu_\mathrm{Drag}$ is a constant within the linear response regime. Therefore, we expect that $\mathrm{Pe}$ is proportional to $\Tilde{F}_\mathrm{ext}$ with proportionality constant $\mu_\mathrm{Drag}/\hat{V}_\mathrm{diff}$ within the linear response regime. This is indeed what we see in Fig.~\ref{fig:app_pefext}.

\section{Stochastic force distribution at $\mathrm{Pe}\approx1.5$}
\label{sec:app_sfpe15}
\begin{figure}[h!]
  \centering
  \includegraphics[width=1.\linewidth]{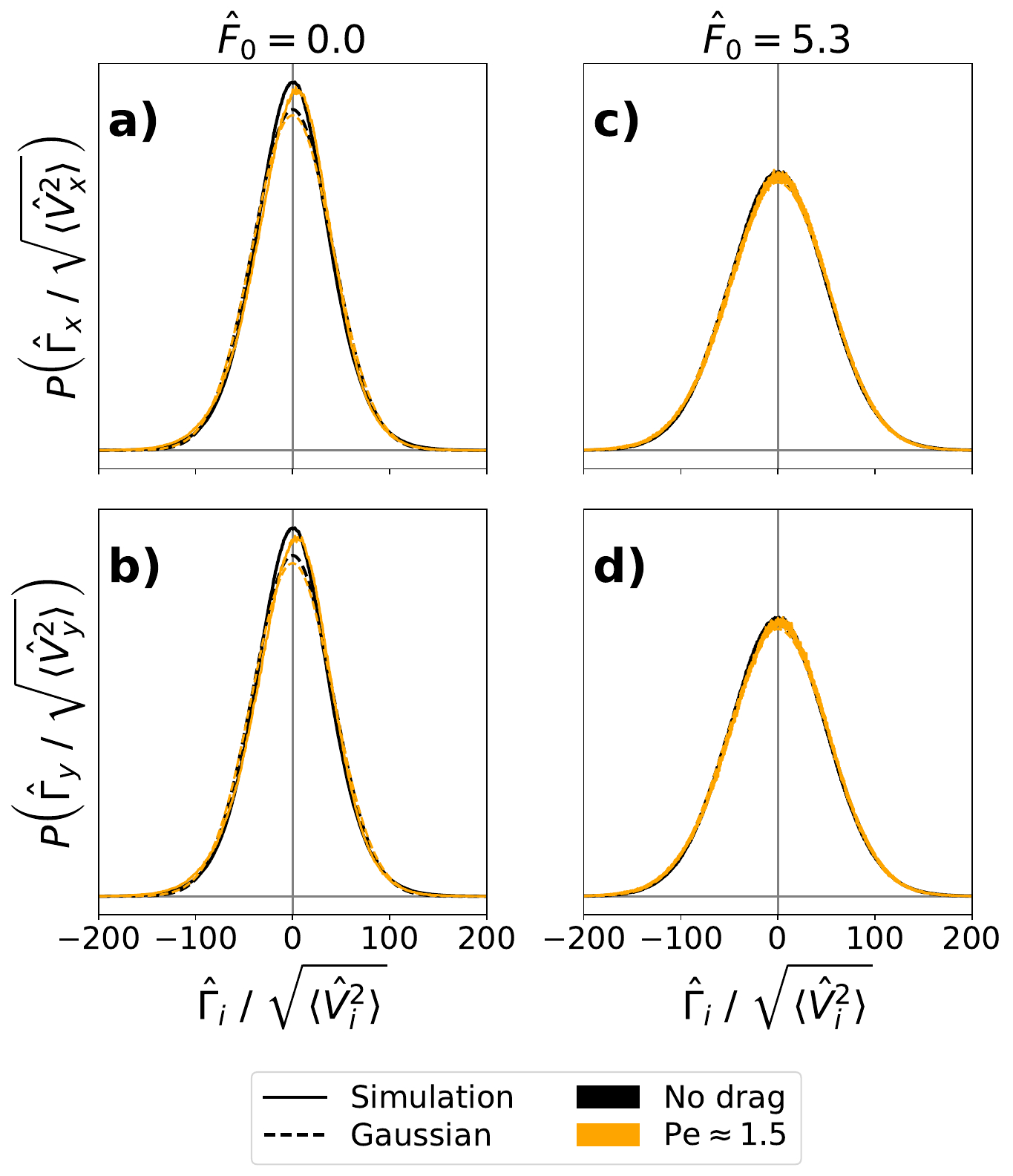}  
\caption{\js{Stochastic force distribution of the dragged, immersed probe in a bath of density $\rho=0.4\sigma^{-3}$. \textbf{a},\textbf{b)} The top row shows the component of the stochastic force which is parallel to the drag force, whereas \textbf{c},\textbf{d)} the bottom row shows one of the perpendicular components (the other perpendicular component is identical). \textbf{a},\textbf{c)} The left column is for a passive bath, whereas \textbf{b},\textbf{d)} the right column is for an active bath ($\hat{F}_0=5.3$). The solid lines show simulation data, whereas the dotted lines show zero-centered Gaussian distributions with the same standard deviation.}}
\label{fig:app_sf}
\end{figure}

\js{In Figs.~\ref{fig:drag_mk}a) and b), we see that the memory kernel of a probe dragged through a passive bath with $\mathrm{Pe}\approx1.5$ already shows slight deviations from that of an undragged probe. Given that these deviations are not as visible in the VACF of the probe in comparison with the memory kernel, we infer that the deviations in the memory kernel were primarily due to the stochastic force distribution on the probe. Indeed, we see in Fig.~\ref{fig:app_sf} that the stochastic force distribution at $\mathrm{Pe}\approx1.5$ in a passive bath already shows a slight asymmetry, a signature of the nonlinear regime, whereas that in an active bath does not. This indicates that the transition to the nonlinear regime in a passive bath occurs more sharply at $\mathrm{Pe}=1$ in comparison with an active bath.}

\bibliography{refs} 

\end{document}